\documentclass[lettersize,journal]{IEEEtran}
\usepackage[flushmargin]{footmisc} 
\usepackage{graphicx}
\usepackage{array}
\usepackage[caption=false,font=normalsize,labelfont=sf,textfont=sf]{subfig}
\usepackage{amsmath}
\usepackage{booktabs}
\usepackage{float}
\usepackage[most]{tcolorbox}
\usepackage{lipsum} 
\usepackage{wasysym}
\usepackage{algorithm}
\usepackage{algorithmicx}
\usepackage{algpseudocode}
\usepackage{amsmath,amsfonts}
\usepackage{svg}
\usepackage{cleveref}
\usepackage{booktabs}   
\usepackage{amssymb}  
\usepackage{textcomp}
\usepackage{stfloats}
\usepackage{url}
\usepackage{verbatim}
\usepackage{graphicx}
\usepackage{cite}

\ifCLASSINFOpdf
\else

\fi

\newcommand{\rot}[1]{\rotatebox{75}{\textbf{\scriptsize #1}}}

\newtcolorbox{promptbox}[1]{
	colback=gray!5!white,      
	colframe=gray!30!black,    
	fonttitle=\bfseries,       
	title={#1},                
	sharp corners,             
	boxrule=1pt,               
	colbacktitle=gray!30!black, 
	coltitle=white,
	enhanced,                  
	attach boxed title to top left={yshift=-2mm, xshift=0mm}, 
}

\begin{document}
\title{GoAT-X: A Graph of Auditing Thoughts for Securing Token Transactions in Cross-Chain Contracts}
\author{\IEEEauthorblockN{Zijun Feng,
		Yuming Feng,
		Yu Wang,
		Weizhe Zhang,~\IEEEmembership{Senior Member,~IEEE}
		Yuhong Nan,
		Yuang Liu and
		Zibin Zheng,~\IEEEmembership{Fellow,~IEEE}}
\thanks{Zijun Feng, Yu Wang, and Weizhe Zhang are with the Harbin Institute of Technology, Shenzhen 518038, China. Zijun Feng, Yuming Feng, and Weizhe Zhang are also with the Peng Cheng Laboratory, Shenzhen 518055, China.
	(E-mail: fengzijun@stu.hit.edu.cn;fengym@pcl.ac.cn;24s151161@stu.hit.edu.cn; 2021111602@stu.hit.edu.cn; wzzhang@hit.edu.cn).}%
\thanks{Yuhong Nan and Zibin Zheng are with Sun Yat-sen University, Guangzhou 510275, China.
	(E-mail: nanyh@mail.sysu.edu.cn; zhzibin@mail.sysu.edu.cn).
}%
\thanks{Yuang Liu is with the Harbin Institute of Technology, Harbin 150001, China.
	(E-mail: 2021111602@stu.hit.edu.cn).}%
\thanks{Weizhe Zhang is the corresponding author.}
}


	
\maketitle

\begin{abstract}
Cross-chain bridges, the critical infrastructure of the multi-chain ecosystem, have become a primary target for attackers, resulting in over \$2.8 billion in losses due to subtle implementation flaws.  Existing defenses, such as bytecode-level static analysis, are ill-equipped to handle the semantic complexity of cross-chain interactions, while LLM-based approaches, which can understand source code, struggle with hallucinatory reasoning over complex, multi-contract dependencies.

In this paper, we propose GoAT-X, a framework that shifts automated cross-chain smart contract codebases auditing from heuristic pattern matching toward systematic first-principles verification. GoAT-X structures the audit process as a Graph of Auditing Thoughts, explicitly mirroring how human experts decompose, reason about, and validate security logic. By anchoring LLM reasoning in statically extracted data flows and explicitly linking abstract security properties to concrete code implementations, the framework constrains semantic reasoning within well-defined structural and state boundaries. Within this constrained space, GoAT-X treats missing constraints and adversarial bypass paths in cross-chain logic as first-class vulnerability targets and dynamically explores reasoning paths to identify exploitable semantic gaps. We evaluate GoAT-X on a comprehensive benchmark covering all known cross-chain token transaction attacks. GoAT-X achieves 92\% recall on fine-grained audit points and 95\% coverage of vulnerable projects, while identifying 117 confirmed risks in the wild with low operational cost, establishing a new standard for scalable, logic-driven cross-chain security.
\end{abstract}

\begin{IEEEkeywords}
Smart Contract Security, Cross-chain Bridge, Graph of Thoughts, Vulnerability Detection, LLM, Static Analysis.
\end{IEEEkeywords}

\section{Introduction}
Cross-chain bridges have become essential infrastructure driven by the demand for DeFi interoperability \cite{ante2021influence, ou2022overview}. However, this essential infrastructure is also highly vulnerable. The complexity of bridge design, stemming from differences in consensus and data structures across chains, introduces novel attack vectors. The typical lock-and-mint mechanism, while functional, is a frequent point of failure \cite{belchior2023hephaestus,lee2023sok}. As their Total Value Locked (TVL) surges, these protocols have become a prime target for attackers, accounting for over \$2.8 billion in losses—nearly 40\% of all Web3 thefts to date \cite{Chainlink2025,slowmist2023blockchain}.

Securing these protocols remains challenging. Manual auditing is unscalable, while automated bytecode analysis suffers from a semantic gap regarding cross-chain intent \cite{liao2024smartaxe, wang2024xguard}.  Although LLM-based approaches offer semantic reasoning\cite{sheng2025llms}, they struggle with cross-chain systems due to their single-chain focus and inability to model multi-contract interactions or generalize to novel architectures\cite{wei2024llm, li2025scalm, liu2024propertygpt, wang2024smartinv}. To the best of our knowledge, currently, there is no effective solution that supports automated, source-code-level auditing of large-scale cross-chain codebases, leaving a critical and unresolved security gap.

Achieving fully automated auditing faces fundamental hurdles due to the intricate and disjointed nature of cross-chain execution, where asynchronous events inherently fragment the continuous control flow required for precise multi-contract modeling. This complexity is further amplified by the sheer scale of cross-chain codebases, which necessitates tracking subtle state dependencies across massive, heterogeneous contract networks to verify security properties effectively.

In this paper, we propose GoAT-X (\textbf{G}raph of \textbf{A}uditing \textbf{T}houghts for \textbf{X}-Chain), a novel framework that shifts cross-chain security from heuristic pattern-based detection to systematic first-principles verification. This approach establishes security by strictly validating the fundamental predicates of Integrity, Authenticity, and Safety against code execution. GoAT-X orchestrates a five-layer automated pipeline that synergizes static analysis, semantic embeddings, and Large Language Models (LLMs) within a structured reasoning graph. Specifically, the framework emulates the cognitive depth of expert auditors by deconstructing the verification task into five progressive stages: locating transaction code, mapping parameters, tracing data flows, verifying constraints, and analyzing bypass paths.

\IEEEpubidadjcol

Crucially, GoAT-X constrains LLM reasoning within explicit structural and state boundaries. Layer 1 and Layer 3 leverage static analysis and context-aware code slicing to extract transaction-specific data flows, ensuring semantic inference is grounded in concrete execution logic. Across all reasoning stages (Layers 2, 4, and 5), we employ model ensembling and a universal self-correction mechanism to rigorously synthesize audit traces while mitigating hallucinations. This pipeline is further refined by encoder-based aggregation in Layer 4 for deduplication, and retrieval-augmented generation (RAG) in Layer 5 to anchor bypass analysis in state context and historical exploit knowledge.

To evaluate GoAT-X, we construct a comprehensive vulnerability benchmark by integrating open-source datasets with incident reports from authoritative security platforms. This dataset comprises 673 smart contracts across 20 distinct cross-chain projects, which we meticulously annotated to establish 294 fine-grained audit points as ground truth. On this benchmark, GoAT-X demonstrates robust performance: it achieves 92\% recall on audit points, reflecting high semantic coverage of security constraints; and successfully achieves 95\% coverage of vulnerable projects in the benchmark, validating its practical detection effectiveness.

\noindent \textbf{Contributions}. We summarize the contributions as follows:
\begin{itemize}
	\item We propose GoAT-X, a framework that reframes vulnerability detection from pattern-based inspection to constrained, first-principles reasoning. By structuring the audit process as a Graph of Auditing Thoughts (GoAT), it synergizes static analysis with LLM reasoning to systematically explore execution paths while minimizing hallucinations.
	\item We define a Cross-Chain Token Transaction Verification Framework that formalizes security properties into enforceable predicates. This addresses the semantic gap in existing tools by providing a rigorous constraint space for identifying missing checks and bypass paths.
	\item  We construct a comprehensive vulnerability benchmark by collecting and deeply annotating a wide range of representative cross-chain token transaction attacks. We then conduct extensive experiments and ablation studies to evaluate the performance and effectiveness of GoAT-X.
\end{itemize}


\section{Preliminary}
\noindent \textbf{Cross-chain Model and Attack Surface.} Smart contracts underpin cross-chain bridges, enabling asset interoperability through a lock-and-mint mechanism. As illustrated in Figure 1, the architecture consists of a source chain, a destination chain, and off-chain relayers, facilitating the following workflow:
\begin{enumerate}
	
	\item Source Chain Operation: A user initiates a request via a Router contract, which triggers the Token contract to lock or burn assets and emits a cross-chain event containing transaction details.
	\item Off-Chain Verification: Off-chain relayers continuously listen to the source chain. Once a cross-chain event is captured, they verify it to ensure the legitimacy of the transaction.
	\item Destination Chain Execution: Upon receiving valid credentials, the destination Router invokes the Token contract to unlock or mint equivalent assets to the user.
	
\end{enumerate}

While this architecture enables interoperability, the disjointed state space introduces novel attack vectors that traditional single-chain security models fail to cover. We categorize the cross-chain attack surface into three semantic layers: \begin{itemize} \item \textbf{Integrity Violation (Malformed Data):} Attackers may initiate transactions with logically invalid parameters (e.g., zero amounts or conflicting token addresses) that bypass superficial checks on the source chain. \item \textbf{Authenticity Violation (Spoofing \& Replay):} The disconnect between chains allows attackers to forge cross-chain messages or replay valid historical messages if cryptographic proofs or nonces are not rigorously enforced. \item \textbf{Safety Violation (Execution Manipulation):} Even with valid messages, the execution context can be manipulated. Attackers may exploit arbitrary external calls or insufficient slippage protection to hijack control flow and drain liquidity. \end{itemize}

\begin{figure}[H]
	\centering
	\includegraphics[width=0.5\textwidth]{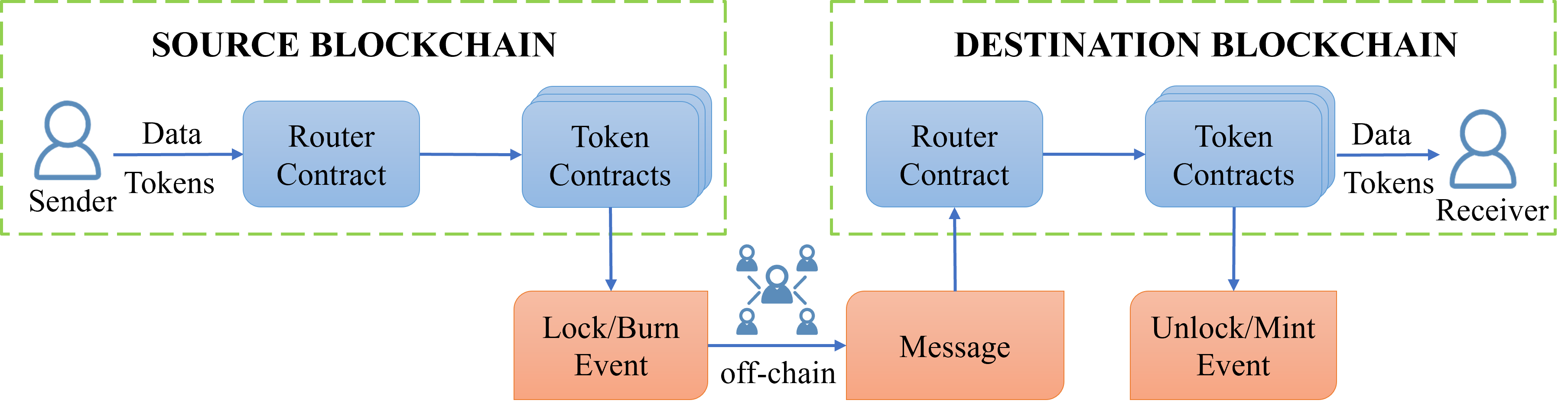}
	\caption{Cross-chain Transaction Workflow}
\end{figure}

\noindent \textbf{Large Language Models (LLMs) and Division of Thoughts.} Large Language Models (LLMs) such as GPT \cite{OpenAIChatGPT2025}, Gemini\cite{GoogleGemini2025}, and Deepseek\cite{DeepSeekChat2025} have demonstrated powerful capabilities in code understanding, generation, and logical reasoning. They are increasingly being used to solve general problems across various tasks. However, during the reasoning process, they are still confined to a character-level, left-to-right decision-making process \cite{yao2023tree}. To address these limitations, researchers have proposed architectures that explore multiple thought processes, such as Tree of Thoughts (ToT) \cite{yao2023tree} and Graph of Thoughts (GoT) \cite{besta2024graph}.

As illustrated in Figure 2, Tree of Thoughts (ToT) structures problem-solving as a search tree, employing branching and backtracking to explore intermediate states and avoid local optima. Graph of Thoughts (GoT) significantly enhances this by transforming the tree into a graph (Figure 2b), introducing novel ``aggregate" and ``refine" operations. These capabilities allow previously separate reasoning paths to converge and share context, enabling a more efficient and controllable exploration of complex reasoning spaces.

Inspired by the ``Division of Thoughts" principle, we apply a similar task decomposition strategy to tackle the complex challenge of auditing cross-chain smart contract codebases.

\begin{figure}[htbp]
	\centering
	\subfloat[Tree of Thoughts]{\includegraphics[width=1.7in]{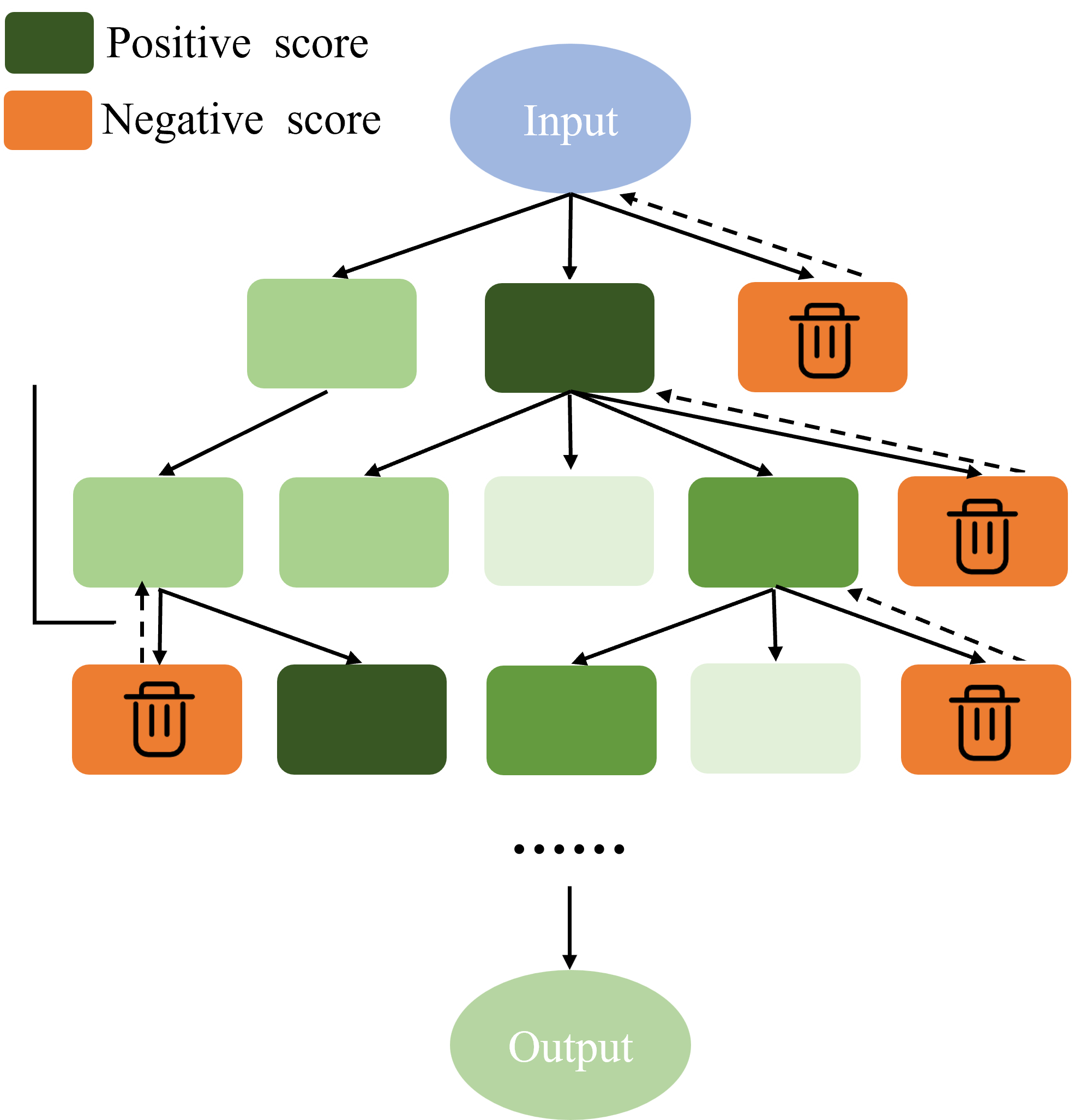}}%
	\label{fig:sub-left}
	\hfill
	\subfloat[Graph of Thoughts]{\includegraphics[width=1.7in]{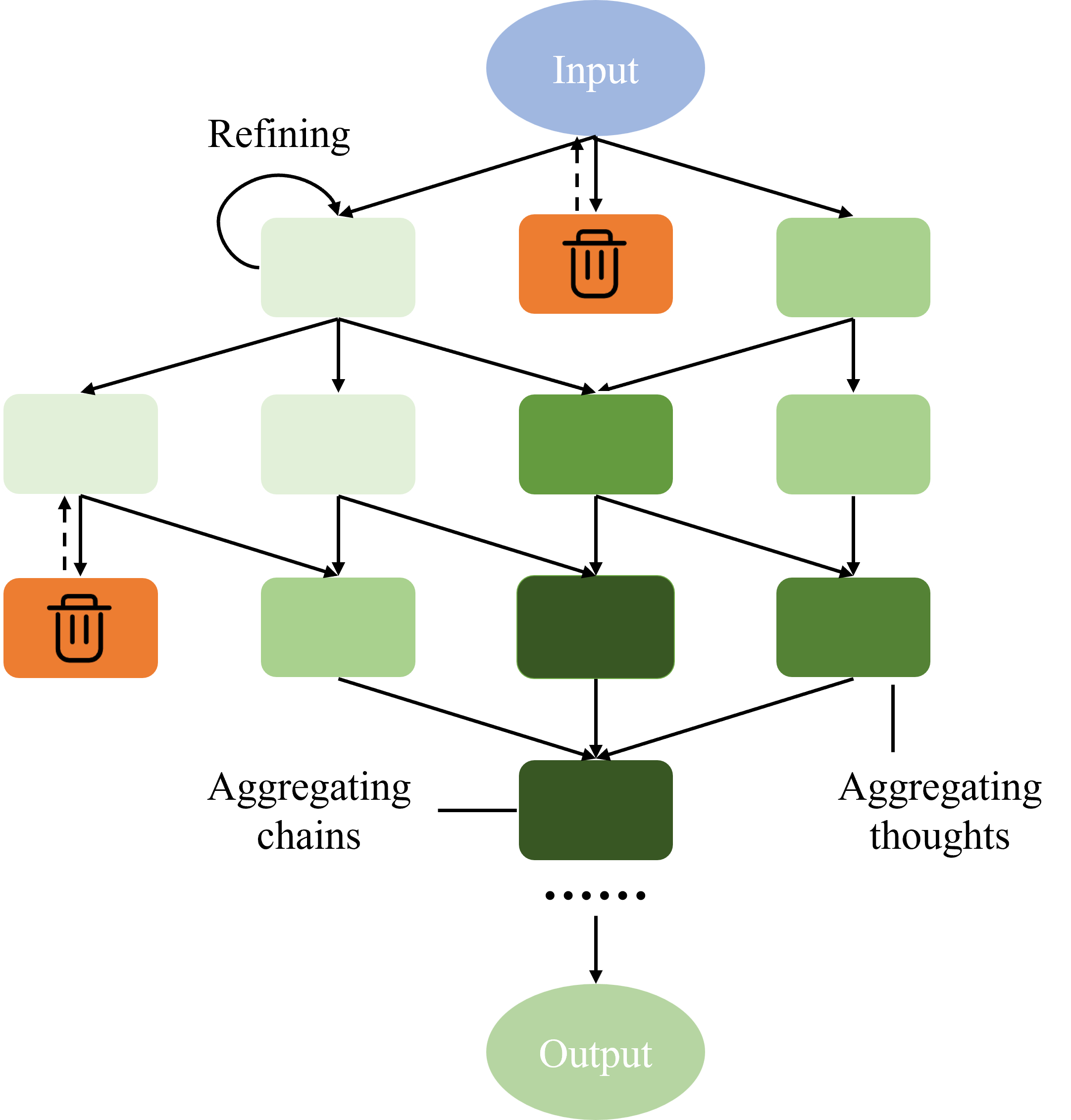}}%
	\label{fig:sub-right}

	\caption{A comparison of thought organization structures. (a) Tree of Thoughts, which explores different reasoning paths in a branching manner. (b) Graph of Thoughts, which allows paths to merge, enabling more complex and efficient exploration.}
	\label{fig:two-subfigs}
\end{figure}

\section{GoAT-X Overview}
\label{sec:GoAT-X}

Through a comprehensive analysis of existing cross-chain vulnerabilities, we observe a unifying security perspective: \textbf{historically observed cross-chain attacks can be attributed to missing constraints or bypassable validations associated with fundamental cross-chain properties}. Specifically, these properties encapsulate the inherent attributes of cross-chain interactions, such as message payloads, external call targets, and price slippage. To defend against varying attack vectors, these properties must be governed by rigorous constraints spanning three core dimensions: \textbf{Integrity}, \textbf{Authenticity}, and \textbf{Safety}. A comprehensive formalization of these properties and their corresponding enforceable constraints is detailed later in \S\ref{sec:verification}. As illustrated in Figure 3, when the constraint of security properties is absent or can be bypassed, adversaries are able to carefully craft cross-chain inputs to violate intended guarantees and maliciously siphon tokens.

This perspective provides the factual foundation for our first-principle verification approach. Guided by it,  we design GoAT-X with the objective of systematically identifying whether security properties in code are enforced by corresponding constraints and whether such constraints can be bypassed in practice. Building on this, we conduct a detailed analysis of the cross-chain execution process and construct a comprehensive Cross-Chain Token Transaction Verification Framework, within which all subsequent reasoning is explicitly constrained. 

Leveraging the event-driven nature of cross-chain logic, GoAT-X uses cross-chain events as anchors to precisely localize and extract the implementation code relevant to each cross-chain transaction from large-scale smart contract codebases. It then abstracts security properties and their associated data flows, and evaluates whether the predicates defined in the Verification Framework are correctly enforced along these data flows or can be bypassed by adversarial executions.

\begin{figure}[H]
	\centering
	\includegraphics[width=0.5\textwidth]{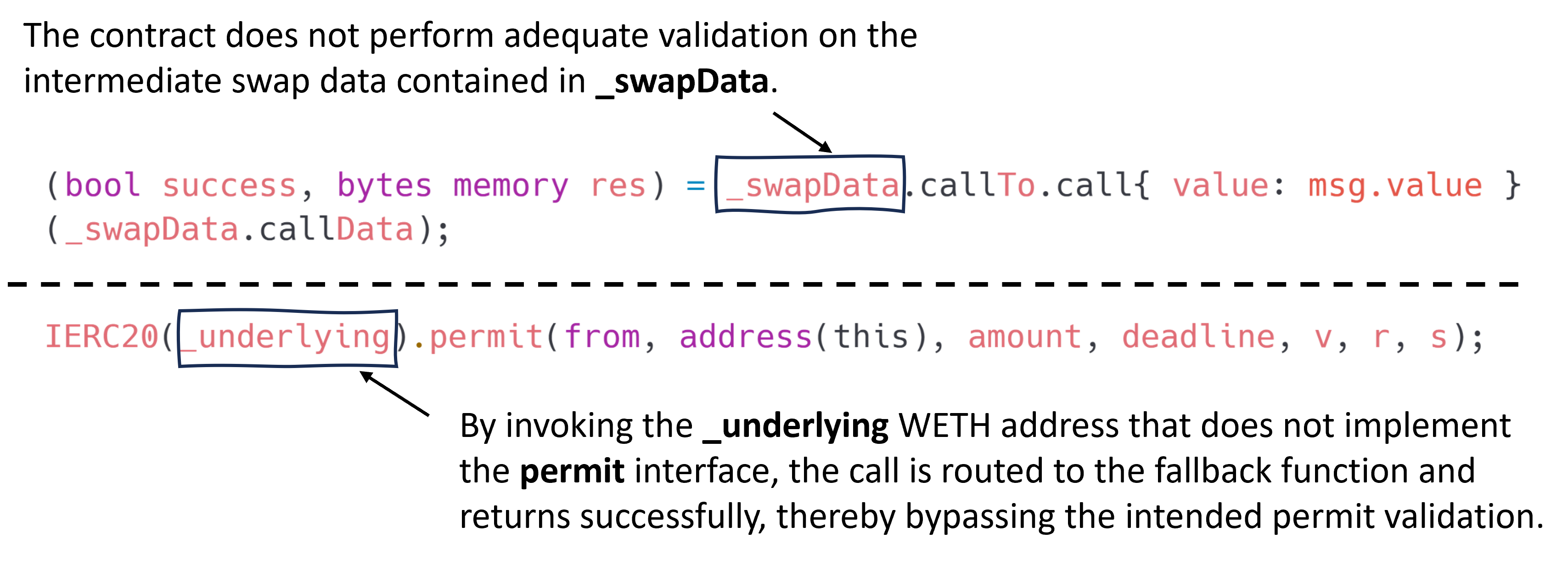}
	\caption{Examples of Missing or Bypassed Security Property Validation}
\end{figure}


As illustrated in Figure 4, GoAT-X adopts a five-layer audit graph architecture. Each node in the graph corresponds to an atomic auditing thought, defined as a structured intermediate reasoning unit generated during the auditing process. Such a thought encapsulates cross-chain contract semantics, partial security properties, associated data flows, constraint implementations, or constraint-bypass inferences. The overall architecture is orchestrated by the Verification Framework Gateway, which manages the flow of auditing thoughts across layers. At each stage, the Gateway normalizes incoming thoughts into structured representations through static analysis and selectively dispatches them to external LLMs or encoder-based components for further reasoning. The resulting outputs are then transformed into atomic auditing thoughts and reintegrated into the audit graph, ensuring that all reasoning remains aligned with the Verification Framework.



By progressively connecting these individual auditing thoughts within the five-layer Graph of Auditing Thoughts, GoAT-X systematically reproduces an expert's entire reasoning chain, culminating in a comprehensive audit report. Implementation details for the Cross-Chain Token Transaction Verification Framework and the Graph of Auditing Thoughts are provided in \S\ref{sec:verification}  and \S\MakeUppercase{\romannumeral 5}, respectively.

\begin{figure*}[hbt]
	\centering
	\includegraphics[width=1.0\textwidth]{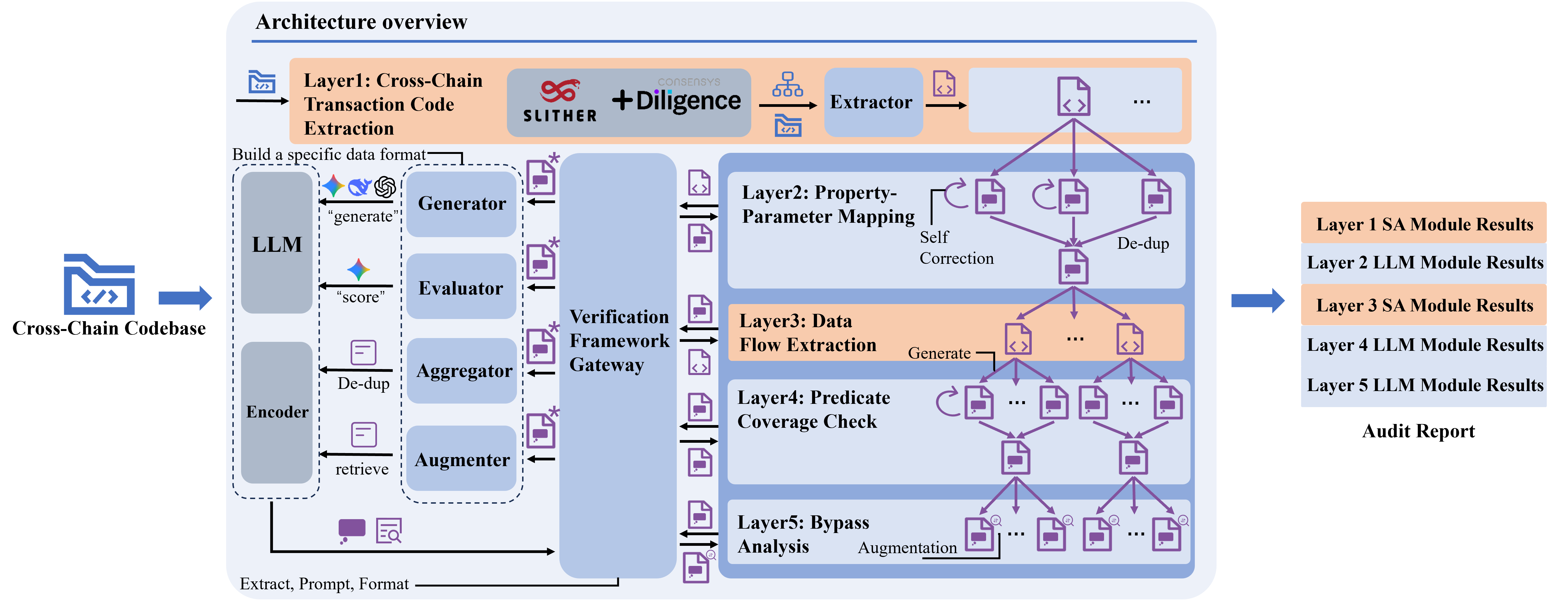}
	\caption{The GoAT-X Architecture. The system integrates a foundational static analysis layer with a multi-stage Large Language Model (LLM) layer to perform comprehensive, end-to-end auditing. The workflow progressively transforms the codebase into a comprehensive Audit Report, composed of structured intermediate results from each layer: extracted transaction code (Layer 1), property-parameter mappings (Layer 2), data flow slices (Layer 3), verified predicates (Layer 4), and confirmed bypass paths (Layer 5).}
\end{figure*}

\section{Cross-Chain Token Transaction Verification Framework}
\label{sec:verification}
We formally define the security properties and corresponding predicates governing the on-chain lifecycle of cross-chain transactions in GoAT-X. These definitions serve as the foundation for the internal Gateway, which applies the underlying principles to systematically guide and constrain the analysis across all layers.

\subsection{Transaction Verification Property Definition}
We define a standardized schema for cross-chain transactions to ensure comprehensive coverage of the attack surface, particularly focusing on externally-controllable parameters and historical vulnerability patterns \cite{zhang2022xscope, liao2024smartaxe, wang2024xguard}.

%
%

First, we define a standardized cross-chain message structure $M$. This structure encapsulates all the core information required from the source chain to the destination chain and serves as the basis for all subsequent on-chain verification logic. The explanation of each field is shown in Table \MakeUppercase{\romannumeral 1}.

\begin{equation}
	\small
	M =\{s_c,s_s,d_c,d_r,t,a,n\}
\end{equation}

\begin{table}[htbp]
	\centering
	\caption{Explanation of Each Field}
	\begin{tabular}{cc}
		\toprule
		\textbf{Field} & \textbf{Explanation} \\
		\midrule
		$s_c$ & Source chain identifier \\
		$s_s$ & Sender address            \\
		$d_c$ & Destination chain identifier \\
		$d_r$ & Receiver address          \\
		$t$   & Token address             \\
		$a$   & Token amount              \\
		$n$   & Nonce                     \\
		\bottomrule
	\end{tabular}
\end{table}

Beyond cross-chain messages, our framework incorporates external contract interactions and slippage parameters (the actual token output, execution price variation, etc.) into the verification scope. This inclusion is necessitated by the prevalence of exploits targeting arbitrary external calls \cite{lifi_exploit_2022} and flash loan attack \cite{halborn_synapse_2021}. Formally, the source-chain property structure $S$ is defined as:

\begin{equation}
	\small
	S =\{M, s_{extAddr}, s_{extFunc}, slippage\}
\end{equation}

For the destination chain contract, verification must strictly target two primary attack vectors: the incoming cross-chain message with its attesting signature, and the external calls executing subsequent business logic. Accordingly, we design the verification property structure as follows:

\begin{equation}
	\small
	D =\{M, d_{extAddr}, d_{extFunc}, signature\}
\end{equation}

\subsection{Transaction Verification Predicates}
To systematically neutralize the attack surface defined in \S\MakeUppercase{\romannumeral 2}, we formalize a set of security predicates grounded in three fundamental principles: Integrity, Authenticity, and Safety. Rather than generic vulnerability patterns, these predicates serve as targeted countermeasures against specific cross-chain threats:
\begin{itemize}
	\item \textbf{Integrity: }Ensures that each user-initiated cross-chain request is complete, unambiguous, and logically well-formed, filtering out malformed or invalid inputs.
	\item  \textbf{Authenticity: } Verifies that a request is genuinely authorized by its claimed originator, is unique, and has not been replayed.
	\item  \textbf{Safety:}  Ensures that executing a valid and authentic request does not introduce unintended risks, including the abuse of the protocol to attack other contracts.
\end{itemize}

To realize the security requirements, we define a comprehensive set of formal predicates at critical checkpoints across the cross-chain lifecycle. (Table \MakeUppercase{\romannumeral 2}) 

\begin{table*}[!htb]
	\centering
	\caption{Cross-Chain Verification Predicate Definitions}
	\begin{tabular}{cc}
		\toprule
		\textbf{Predicate} & \textbf{Explanation} \\
		\midrule
		$ChainId()$ & The chain ID of the current contractr \\
		$ZeroAddr$ & Zero address            \\    
		$SupportedChains$ & The set of chains supported by the protocol \\
		$Balance(address,token)$ & An address's token balance          \\
		$NonceNext(sender)$   & The source chain's next expected nonce for each sender             \\
		$NonceUnused(chain,nonce)$   & The nonce on the destination chain has not been used yet              \\
		$AddrWhitelist,FuncWhitelist$   &  Address/Function Whitelist                     \\
		$ValidProof(M,signature)$   & Verify the integrity and correctness of the signature \\
		$AssetMap(chain,token)$  & Secure asset registry on the chain \\
		$LockedCorrect(sender,token,amount)$  & The balance changes correctly after the source chain contract assets are locked/burnt \\
		$UnlockedCorrect(receiver,token,amount)$ & The balance changes correctly after the destination chain assets are unlocked/minted \\
		$MinExecutionBounded(slippage)$ & The user’s received amount is bounded by a minimum execution threshold \\
		$ReferencePriceBounded(slippage)$ & The execution price deviation is bounded against a protocol-defined or external reference price \\
		\bottomrule
	\end{tabular}
\end{table*}

These predicates collectively form the Verification Framework, in which $P_s$ represents the source chain predicate logic and $P_d$ represents the destination chain predicate logic:

\noindent \textbf{Integrity Verification:} 

\begin{equation}
	\small
	\begin{split}
		P_s^{(1)}(S):=&d_r \ne ZeroAddr \land a>0  \land \\
		&d_c \ne ChainId()
	\end{split}
\end{equation}

\begin{equation}
	\small
	\begin{split}
		P_d^{(1)}(D):=&d_r \ne ZeroAddr \land a>0 \land \\
		& d_c = ChainId()
	\end{split}
\end{equation}

These verification steps formalize parameter sanity checks to exclude malformed requests and ensure logical soundness. The pNetwork bridge exploit \cite{zhang2024attack} illustrates the consequences of missing such validation, where zero-amount transactions were accepted.

\noindent \textbf{Authenticity Verification:} 

\begin{equation}
	\small
	\begin{split}
		P_s^{(2)}(S):=&t \in AddrWhitelist \land \\
		&n = NonceNext(s_s) \land \\
		&d_c \in SupportedChains 
	\end{split}
\end{equation}

\begin{equation}
	\small
	\begin{split}
		P_d^{(2)}(D):=&NonceUnsed(s_c, n) \land \\
		&s_c \in SupportedChains \land \\
		&ValidProof(M,signature)  
	\end{split}
\end{equation}

Authenticity verification formalizes message authorization and temporal uniqueness, ensuring that each request is legitimately authorized and executed exactly once. Failures in this layer, such as the Nomad bridge exploit \cite{wu2025safeguarding}, demonstrate how flawed authorization logic can lead to arbitrary message execution. 

\noindent \textbf{Safety Verification:}

\begin{equation}
	\small
	\begin{split}
		P_s^{(3)}(S):=&s_{extAddr} \in AddrWhitelist \land \\
		&s_{extFunc}\in FuncWhitelist \land \\ 
		& LockedCorrect(s_s,t,a) \land \\
		&MinExecutionBounded(slippage) \land  \\
		&ReferencePriceBounded(slippage) 
	\end{split}
\end{equation}

\begin{equation}
	\small
	\begin{split}
		P_d^{(3)}(D):=&d_{extAddr} \in AddrWhitelist \land \\
		&d_{extFunc}\in FuncWhitelist \land \\ 
		& UnLockedCorrect(s_s,t',a)
	\end{split}
\end{equation}

Safety verification formalizes execution-time constraints to prevent unintended side effects, ensuring that legitimate requests cannot be abused to compromise the ecosystem. Lapses in such checks, like the vulnerability in Qbridge that failed to validate account locks \cite{ODAILY_QubitQBridge_2022}, result in significant financial damage.

The final execution condition is that the source chain contract executes successfully if and only if:
\begin{equation}
	\small
	P_s^{(1)}(S) \land P_s^{(2)}(S) \land P_s^{(3)}(S)
\end{equation}

The destination chain contract execution is successful if and only if:
\begin{equation}
	\small
	P_d^{(1)}(D) \land P_d^{(2)}(D) \land P_d^{(3)}(D)
\end{equation}

Together, these predicates constitute a defense-in-depth verification logic that precisely delineates the cross-chain attack surface and enables systematic detection of authorization, validation, and execution flaws. This logic serves as the formal backbone of GoAT-X, supporting sound, extensible, and automated cross-chain security analysis.

\section{The Five Auditing Layers of the GoAT-X}
\label{sec:five layer}
%

We aim to perform a deep, end-to-end analysis of cross-chain codebases to ensure that non-bypassable validation policies, as prescribed by our Verification Framework, are correctly implemented at the code level. To tackle this complex task, we emulate an expert's audit methodology by deconstructing the verification task into five manageable sub-tasks, each corresponding to a layer in our architecture: locating code, identifying parameters, tracing data flows, verifying constraints, and analyzing bypasses. Together, these layers construct a structured Graph of Auditing Thoughts.

Different auditing tasks require different analytical capabilities. We therefore combine two complementary modules: static analysis for precise and complete data-flow reasoning, and LLM-based analysis for semantic interpretation and abstraction. This hybrid design enables efficient and accurate static analysis, complemented by large language models that resolve semantic ambiguities beyond the reach of traditional techniques. 

Formally, we model our auditing process as a tuple $(G,\mathcal{O})$, where $G=(S,E)$ is a graph representing the reasoning process, and $\mathcal{O}=\{\mathcal{G}, \mathcal{T}, \mathcal{V}, \mathcal{R}\}$ denotes a set of operations that act on this graph. Lowercase symbols (e.g., $x, y, z$) denote atomic auditing thoughts generated during the analysis. Within the graph $G$, each node is a state $s\in S$, defined as $s=[x,z_{1...i}]$. This state encapsulates the initial input $x$ and the sequence of thoughts $z_{1...i}$ generated thus far. $E\subseteq S \times S $ is a set of directed edges, representing the transitions between reasoning states. This formalization provides a unifying abstraction for integrating static analysis and LLM-driven reasoning within a single audit graph.

\noindent \textbf{Static Analysis (SA) Module:}
Serving as the core engine for Layer 1 and Layer 3, this module leverages external tools to construct Abstract Syntax Trees (ASTs) for cross-chain contract repositories. Based on the ASTs, it identifies cross-chain token transaction code and performs data-flow analysis to extract security properties within code snippets. We denote the static analysis module as $q$ and define the operation to obtain structured thoughts based on the current state $s$ as $R(q, s)$. This operation yields a variable-sized set of thoughts  based on the distinct logical implementations within the code:

\begin{equation}
	\mathcal{R}(q, s) \sim q([ z_i^{1...k_s}] \mid s)
\end{equation}

where the number of thoughts $k_s$ is determined by the distinct cross-chain logics identified in the code.

\noindent \textbf{Large Language Model (LLM) Module:}
Serving as the core engine for Layers 2, 4, and 5, the LLM module adopts a multi-stage generate–aggregate–evaluate paradigm to conduct deep security analysis. This structured approach improves result correctness through cross-validation, enabling reasoning capabilities that extend beyond single-pass inference. At each layer, GoAT-X employs an ensemble of different LLMs queried with structured prompts to generate multiple candidate reasoning paths. These paths are systematically merged into a unified thought, which is then evaluated. The most reliable reasoning outcome is selected based on confidence scoring. To improve robustness, we incorporate self-correction to suppress hallucinated inferences and retrieval-augmented generation (RAG) to ground high-level bypass analysis in a protocol-specific context.

We denote the LLM module parameterized by $\theta$ as $p_\theta$, and define the following core operations:
\begin{enumerate}
	\item \textbf{Thoughts Generation ($\mathcal{G}$)}:This operation expands the reasoning space from a given state $s$ by leveraging an ensemble of $k_\mathcal{G}$ LLMs to generate diverse parallel thought candidates. Each model produces next-step thoughts conditioned on the current state.
	\begin{equation}
		\mathcal{G}(p_\theta,s,k_G) \sim p_{\theta}^{generate}([z_{i+1}^{1...k_\mathcal{G}}]|s)
	\end{equation}
	\item \textbf{State Aggregation ($\mathcal{T}$)}:This operation consolidates a set of candidate states $S$ by merging semantically or structurally equivalent thoughts. It employs encoder-based similarity scoring and rule-based matching to prune redundant outputs and to merge thoughts corresponding to the same security property or constraint into a unified thought $z_{i+1}^{\prime}$:
	\begin{equation}
		\mathcal{T}(p_{\theta}, S)\sim p_{\theta}^{aggregate}(z_{i+1}^{\prime}|S)
	\end{equation}
	\item \textbf{State Evaluation ($\mathcal{V}$)}: This operation evaluates candidate property or constraint analyses $a$ within a given state and selects the highest-scoring one to form a consolidated thought. An LLM-based value function $p_\theta^{\text{value}}$ assigns confidence scores $v \in [0,100]$ to each candidate, and the top-ranked result is retained as the output thought.
	\begin{equation}
		\mathcal{V}(p_\theta,s) \sim \operatorname*{argmax}_{a \in s} {p_{\theta}^{value}(v|s, a)}
	\end{equation}
\end{enumerate}

The overall methodology integrating these operations is detailed in the pseudocode in Algorithm 1. While the algorithm generates sets of state nodes $S_t$ at each layer, the edges $E_t$ are implicitly defined by the parent-child relationships. By formalizing these connections, for instance as $E_t=\{(parent(s),s)| s\in S_t\}$, we can then proceed to construct the five-layer auditing graph $G=(\cup_{t=1}^5S_t,\cup_{t=1}^5E_t)$. 

\begin{algorithm}[H]
	\small
	\caption{Exploration of Auditing Layers}
	\label{alg:got_framework}
	\begin{algorithmic}[1]
		\Statex \textbf{Require:} codebase $C$, size of the model ensemble $k_\mathcal{G}$
		
		\State $S_0 \gets \{[C]\}$
		\For{t=1,...,5}
		\State $S_t \gets \emptyset$ 
		\For{$s$ in $S_{t-1}$}
		\If{\text{layer t in LLM module}}
		\State $S_{t}^{\prime} \gets \{[s,z_t]| z_t\in \mathcal{G}(p_\theta,s,k_\mathcal{G})\}$
		\State $s^{\prime} \gets [s,\mathcal{T}(p_\theta, S_t^{\prime})]$
		\State $S_t \gets S_t \cup \{[s,\mathcal{V}(p_\theta,s^{\prime})]\}$
		
		\EndIf
		\If{\text{layer t in SA module}}
		\State $S_t \gets S_t \cup \{[s,z_t]| z_t\in \mathcal{R}(q,s)\}$
		\EndIf
		\EndFor
		\EndFor
		\State \textbf{return} $\cup_{t=1}^5S_t$ 
		
	\end{algorithmic}
\end{algorithm}

\subsection{Layer 1: Cross-Chain Transaction Code Extraction(SA)} The primary objective of this layer is to decompose a large and complex codebase into a set of foundational auditing thoughts, each capturing a self-contained and transaction-specific code flow. By constraining the analysis to these critical flows, the layer enables fine-grained inspection while effectively focusing the LLM on the most relevant implementation logic.

To this end, GoAT-X first constructs a directed function call graph for the entire codebase using Slither \cite{feist2019slither} and Surya \cite{ConsenSysDiligence_Surya_2025}, as illustrated in Figure 5. Functions emitting key cross-chain events are then identified as anchors. Starting from these anchors, a dual Breadth-First Search (BFS) traversal is performed to isolate the complete execution logic of each cross-chain transaction.

\begin{figure}[H]
	\centering
	\includegraphics[width=0.5\textwidth]{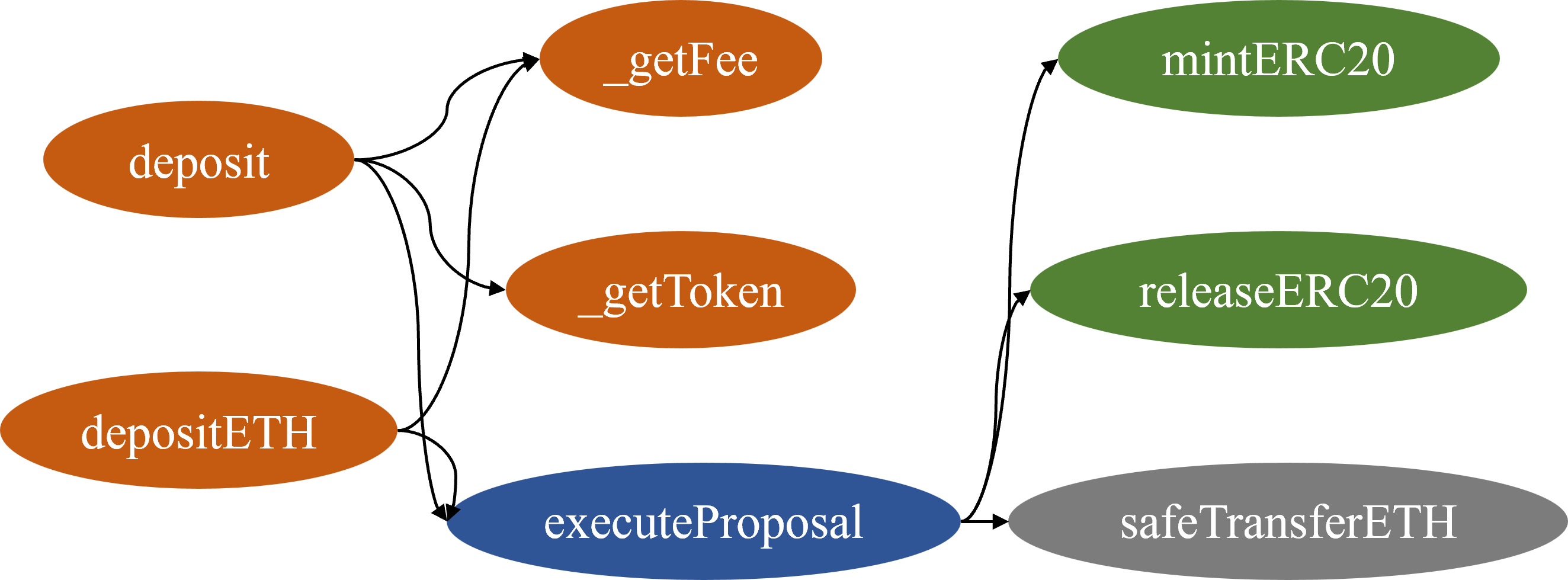}
	\caption{A partial illustration of the function call graph. Nodes of different colors belong to different smart contracts, clearly showing the interactions between them.}
\end{figure}

Guided by event anchors on both the source and destination chains, this dual-traversal strategy precisely localizes the code relevant to a cross-chain operation. Each extracted flow is treated as a node in the auditing graph and serves as the trusted input for all subsequent analysis layers (Algorithm 2).

\begin{algorithm}[htbp]
	\small
	\caption{Extracting Cross-Chain Functions as Nodes}
	\label{alg:extraction}
	\begin{algorithmic}[1]
		\Statex \textbf{Require:} codebase $C$, cross-chain event $Event$
		\State $G \gets \Call{BuildGraph}{C}$ \Comment{Using Slither \& Surya}
		\State $E \gets \Call{FindFunction}{C, Event}$ \Comment{Regex for events}
		\State $N \gets \emptyset$ 
		
		\For{each function e in E}
		\State $S \gets \{e\}$  \Comment{Backward BFS}
		\State $S^{\prime} \gets \{e\}$
		\While{$S^{\prime}$ is not empty}
		\State $S^{\prime} \gets \{u\mid u\notin S,\; v\in S,\; edge(u,v)\in  G\} $
		\State $S \gets S^{\prime} \cup S$
		\EndWhile
		\State $S \gets \{s|s\in S,s \text{ has } \texttt{"public"} \text{ or } \texttt{"external"} \}$
		\Statex
		\For{$s \in S$}: \Comment{Forward BFS}
		\State $n \gets \{s\}$
		\State $S^{\prime} \gets \{s\}$
		\While{$S^{\prime}$ is not empty}
		\State $S^{\prime} \gets \{v\mid u\in n,\; v\notin n,\;edge(u,v)\in G\} $
		\State $n \gets S^{\prime} \cup n$
		\EndWhile
		\State Add $n$ to $N$
		\EndFor
		\EndFor
		
		\State \Return \textit{N}
		
	\end{algorithmic}
\end{algorithm}

\subsection{Layer 2: Property-Parameter Mapping(LLM)} We formulate this step as an entity recognition task, where large language models are leveraged to establish high-confidence semantic mappings between abstract security properties and concrete code parameters. Figure 6 illustrates the overall workflow.

Given an analysis node from Layer 1, the Gateway enriches it with relevant security properties and forwards a unified code–property context to the Generator. The Generator issues a single consolidated prompt to an ensemble of 
$k_\mathcal{G}$ LLMs, enabling diverse yet complementary property–parameter hypotheses.

The resulting mappings are aggregated to remove redundancy and then validated by an Evaluator, which assigns confidence scores to each candidate. For each security property, only the highest-scoring mapping is retained, while low-confidence mappings and nodes are pruned, preventing conflicting semantic bindings and enforcing a deterministic, high-precision output.

Each validated property–parameter mapping is finally encapsulated as a new auditing thought node, forming the output of this layer. Prompt templates are provided in Appendix.

\begin{figure}[H]
	\centering
	\includegraphics[width=0.5\textwidth]{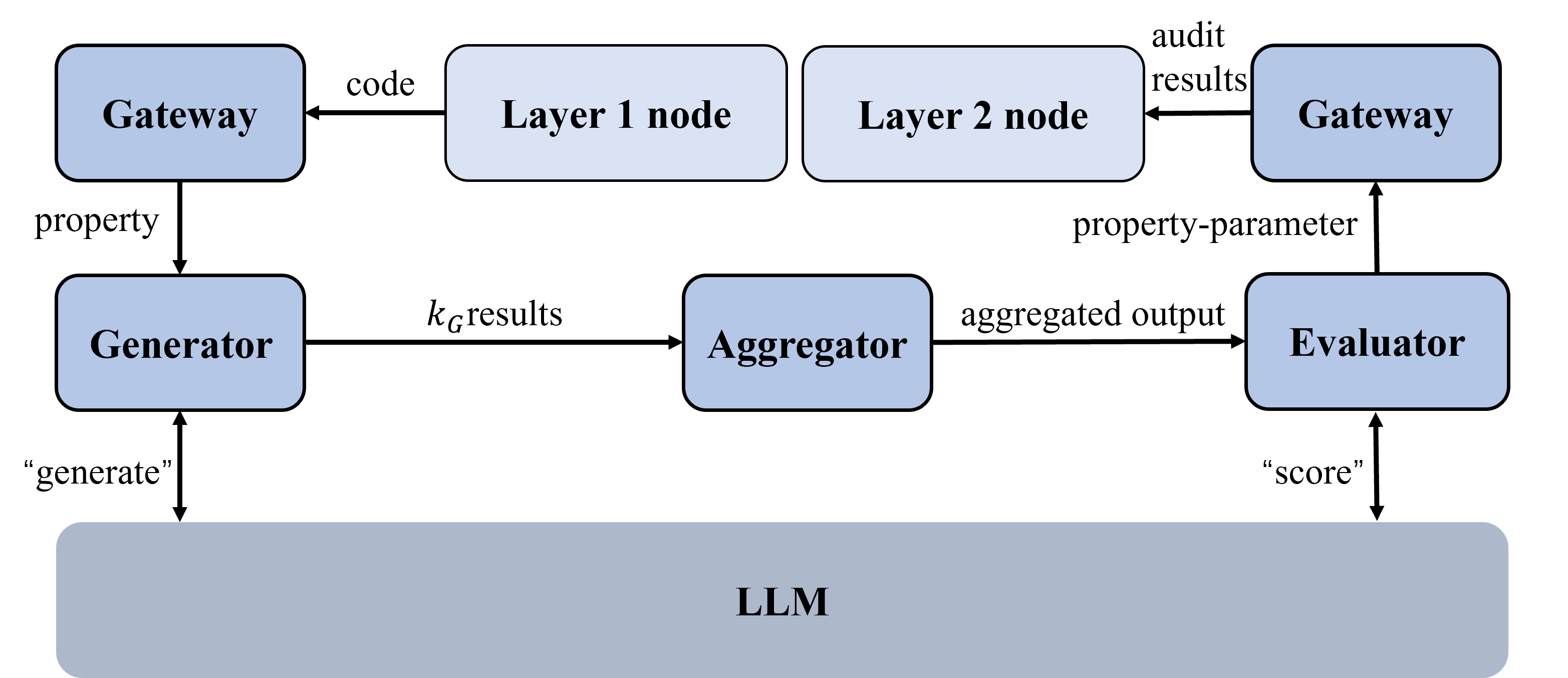}
	\caption{Workflow of Layer 2}
\end{figure}

\subsection{Layer 3: Data Flow Extraction(SA)} To isolate parameter-specific data flow slices and provide high-signal inputs for subsequent LLM reasoning, we design a lightweight static data flow analyzer. While LLM-based methods\cite{wang2024llmdfa} can perform sliced dataflow extraction, they typically incur prohibitive computational costs and latency while remaining prone to hallucinations. Leveraging the extracted parameters and the codebase AST constructed in Layer 1, the analyzer identifies all parameter-related tainted variables $\mathcal{T}$ through a three-stage process.
\begin{enumerate}
	\item \textbf{Bidirectional Taint Propagation:} Unlike traditional unidirectional analysis, we adopt an aggressive infection strategy to capture implicit dependencies. For any statement $\sigma$, let $\mathcal{V}_{\sigma}$ be the set of all identifiers involved. The taint set $\mathcal{T}$ is updated as:
	\begin{equation}
		\mathcal{T}_{next} = \mathcal{T} \cup { \mathcal{V}_{\sigma} \mid \mathcal{V}_{\sigma} \cap \mathcal{T} \neq \emptyset }
	\end{equation}
	\item \textbf{Inter-procedural Mapping:} To handle cross-contract logic, the analyzer maintains a function signature cache $\mathcal{S}: \text{func} \rightarrow \{p_1, \dots, p_n\}$. Upon encountering a FunctionCall at a call site with arguments $\{a_1, \dots, a_n\}$, the state is propagated via:
	\begin{equation}
		a_i \in \mathcal{T} \implies p_i \in \mathcal{T}
	\end{equation}
	\item \textbf{Convergence:} The module iteratively scans the AST until $|\mathcal{T}|$ reaches a fixed point.
\end{enumerate}

After identifying the taint set $\mathcal{T}$, the module applies context-aware code slicing to construct a concise data flow representation $Z_{SA}$. To materialize the sliced AST into a form suitable for downstream reasoning, We define a recursive operator $\text{Rec}: \mathcal{N} \rightarrow \mathcal{S}$ to map AST nodes to their corresponding Solidity code representations. A partial specification is given as follows:

\begin{equation}
	\small
	\text{Rec}(\sigma) =
	\begin{cases}
		\text{name}(\sigma) & \sigma \in \text{Identifier} \\
		\text{Rec}(\sigma_{left}) \oplus \text{op} \oplus \text{Rec}(\sigma_{right}) & \sigma \in \text{BinaryOp} \\
		\text{Rec}(\sigma_{head}) \cup \text{Rec}(\sigma_{body}) & \sigma \in \text{FuncDef}
	\end{cases}
\end{equation}

To preserve path constraints essential for security reasoning, the slicer forcibly includes enclosing branch conditions whenever a tainted statement appears within a conditional block, even if the condition itself is not directly tainted. This design ensures that the resulting slices retain sufficient semantic context for constraint verification while significantly reducing token complexity. Algorithm 3 presents a simplified Context-Aware Code Slicing procedure, which generates parameter-specific data flow slices for each target variable. This layer deliberately favors conservative static slicing over expressive but unstable LLM-based extraction, trading completeness for robustness and predictable cost.

\begin{algorithm}[t]
	\small 
	\caption{Context-Aware Code Slicing}
	\label{alg:slicing}
	\begin{algorithmic}[1]
		\renewcommand{\algorithmicrequire}{\textbf{Input:}}
		\renewcommand{\algorithmicensure}{\textbf{Output:}}
		\State \textbf{Require:} AST Map $\mathcal{A}$, Taint Set $\mathcal{T}$
		\Function{Extract}{$node, \mathcal{C}, force$}
		\State $is\_tainted \gets \text{CheckTaint}(node, \mathcal{T})$
		\State $z_{local} \gets \emptyset$
		
		\If{$node.type \in \{\text{If, For, While}\}$}
		\State $ctx \gets \text{Rec}(node.condition)$
		\State $\mathcal{C}^\prime \gets \mathcal{C} \cup \{ctx\}$
		\State $z_{sub} \gets \Call{Extract}{node.body, \mathcal{C}^\prime, is\_tainted \lor force}$
		
		\If{$z_{sub} \neq \emptyset \lor is\_tainted \lor force$}
		\State $z_{local} \gets z_{local} \cup \{(ctx, \mathcal{C})\}$
		\State $z_{local} \gets z_{local} \cup z_{sub}$
		\EndIf
		
		\ElsIf{$node.type = \text{Block}$}
		\For{\textbf{each} $stmt \in node.statements$}
		\State $z_{local} \gets z_{local} \cup \Call{Extract}{stmt, \mathcal{C}, force}$
		\EndFor
		
		\ElsIf{$is\_tainted \lor force$}
		\State $code \gets \text{Rec}(node)$
		\State $z_{local} \gets z_{local} \cup \{(code, \mathcal{C})\}$
		\EndIf
		
		\State \Return $z_{local}$
		\EndFunction
		
		\State $Z_{raw} \gets \emptyset$
		\For{\textbf{each} $root \in \mathcal{A}$}
		\State $Z_{raw} \gets Z_{raw} \cup \Call{Extract}{root, \emptyset, \text{False}}$
		\EndFor
		
		\State $Z_{SA} \gets \text{Reformat}(Z_{raw})$ \Comment{Refactor into standard code format}
		\State \Return $Z_{SA}$
	\end{algorithmic}
\end{algorithm}

\subsection{Layer 4: Predicate Coverage Check(LLM)}

Leveraging the insight that LLMs reason more effectively over natural language than symbolic logic \cite{xu2024faithful}, the Gateway first translates each formal security predicate into a descriptive, natural language format. 

The Generator then produces $k_\mathcal{G}$ candidate analyses, identifying potential constraint implementations within the parameter-specific data flow slice. To avoid redundant reasoning, the Aggregator performs semantic-level pruning across these LLM-generated candidates by comparing the semantic similarity of their constraint-implementing code snippets.

Specifically, each snippet is encoded using UniXcoder\cite{guo2022unixcoder}, a pretrained model with strong cross-language generalization for semantic code representation and clone detection, to obtain a contextual embedding $\mathbf{v}$, followed by $L2$
normalization. We quantify the redundancy between candidates $A$ and $B$ using cosine similarity, which reduces to a dot product for normalized embeddings:
\begin{equation}
	Similarity(A, B) = \mathbf{v}_A \cdot \mathbf{v}_B
\end{equation}

Candidates exceeding a similarity threshold $k_\mathbf{v}$ are pruned, retaining only unique semantic logic.  This pruning stage serves as a low-cost filter, significantly reducing token consumption by preventing the expensive Evaluator from processing semantically equivalent analyses. 

Finally, the remaining candidates are scored for correctness, and the highest-confidence verified implementation is forwarded to the bypass analysis stage. The overall workflow is illustrated in Figure 7, with detailed prompt templates provided in Appendix.

\begin{figure}[H]
	\centering
	\includegraphics[width=0.5\textwidth]{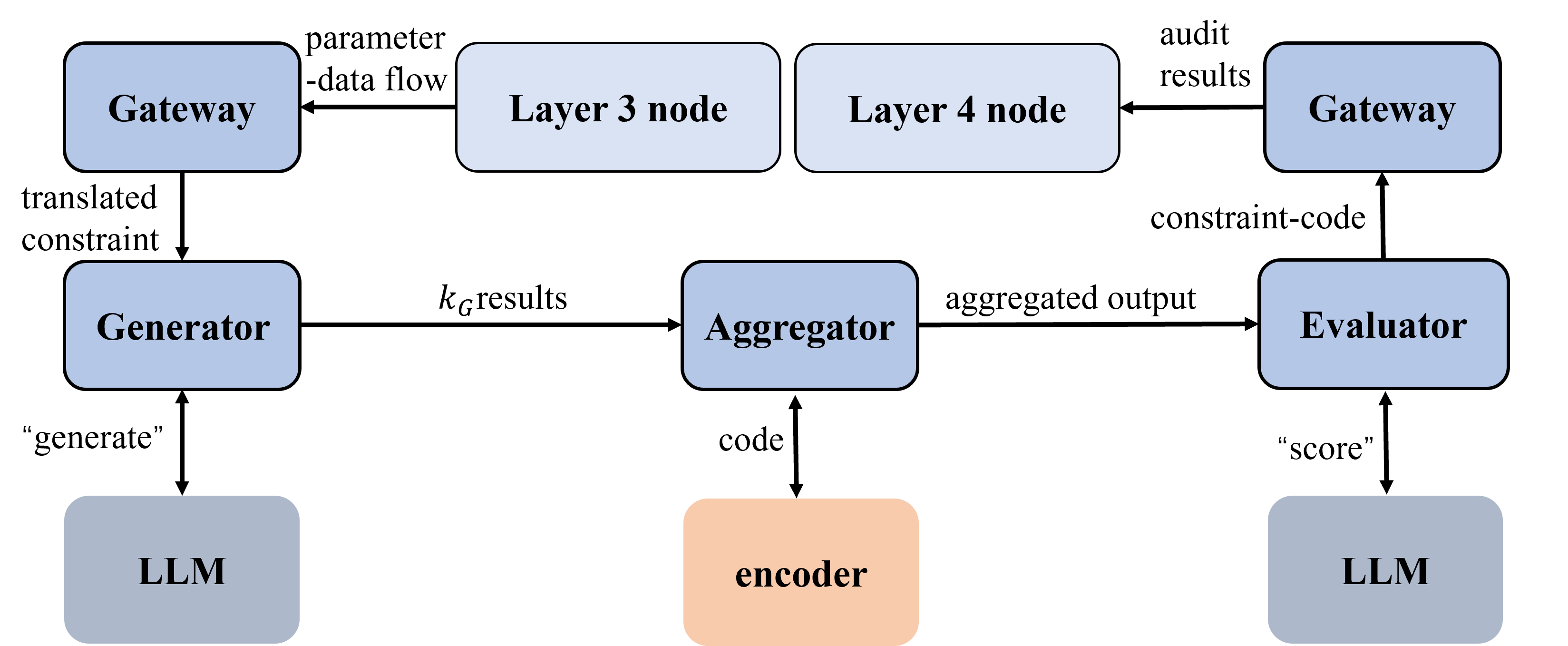}
	\caption{Workflow of Layer 4}
\end{figure}

\subsection{Layer 5: Bypass Analysis(LLM)}
Historical incident analysis shows that verification bypasses often stem from two overlooked vectors: improper configuration of state variables and pre-existing single-chain vulnerabilities. To address these challenges, Layer 5 adopts an attacker-centric reasoning strategy that augments LLM analysis along two complementary dimensions: state-aware execution context and experience-driven bypass knowledge. The Augmenter incorporates two auxiliary retrieval-augmented generation (RAG) modules to systematically guide bypass exploration.

\noindent \textbf{State Context Enhancement}. The module enriches the analysis with precise state information (Figure 9). Each function in the codebase is assigned a stable identifier derived from its original source, enabling efficient retrieval of its associated state variables and initialized values from the imported contracts. During analysis, the recovered state context is injected into the prompt, allowing the LLM to reason about bypasses that depend on concrete initialization or runtime state, rather than code structure alone.

\noindent \textbf{Bypass Knowledge Retrieval}. To further guide attacker-style reasoning, this module supplies principle-level bypass knowledge distilled from real-world exploits. We curate representative constraint-bypass patterns from established sources, including the SWC registry~\cite{swc_registry} and the 101 DeFi Hacks dataset~\cite{defi_hack_labs}. For efficient similarity-based retrieval, representative code snippets are encoded using UniXcoder to obtain semantic embeddings, which are indexed in a FAISS vector database using inner-product similarity. Instead of exposing exploit-specific code, each pattern is abstracted into structured metadata that summarizes its underlying bypass principle and root cause. During analysis, retrieved bypass principles are provided as few-shot references, steering the LLM toward historically effective attack strategies while mitigating hallucinated or implausible reasoning.

After the Gateway passes the intermediate outputs from previous layers, the Augmenter enriches the analysis context with state-aware and bypass-oriented information. The augmented context from both modules instructs the Generator to act as an attacker and produce $k_\mathcal{G}$ independent analyses, each proposing a potential bypass path along with a viable Proof of Concept (PoC). The Aggregator merges valid outputs and removes empty responses. Each remaining bypass candidate is then passed to the Evaluator for a rigorous second-level review and confidence scoring. Only scenarios exceeding a predefined threshold are confirmed as valid vulnerabilities and included in the final report. The overall workflow is illustrated in Figure 8, with detailed prompt templates provided in Appendix.
\begin{figure}[t!]
	\centering
	\includegraphics[width=0.5\textwidth]{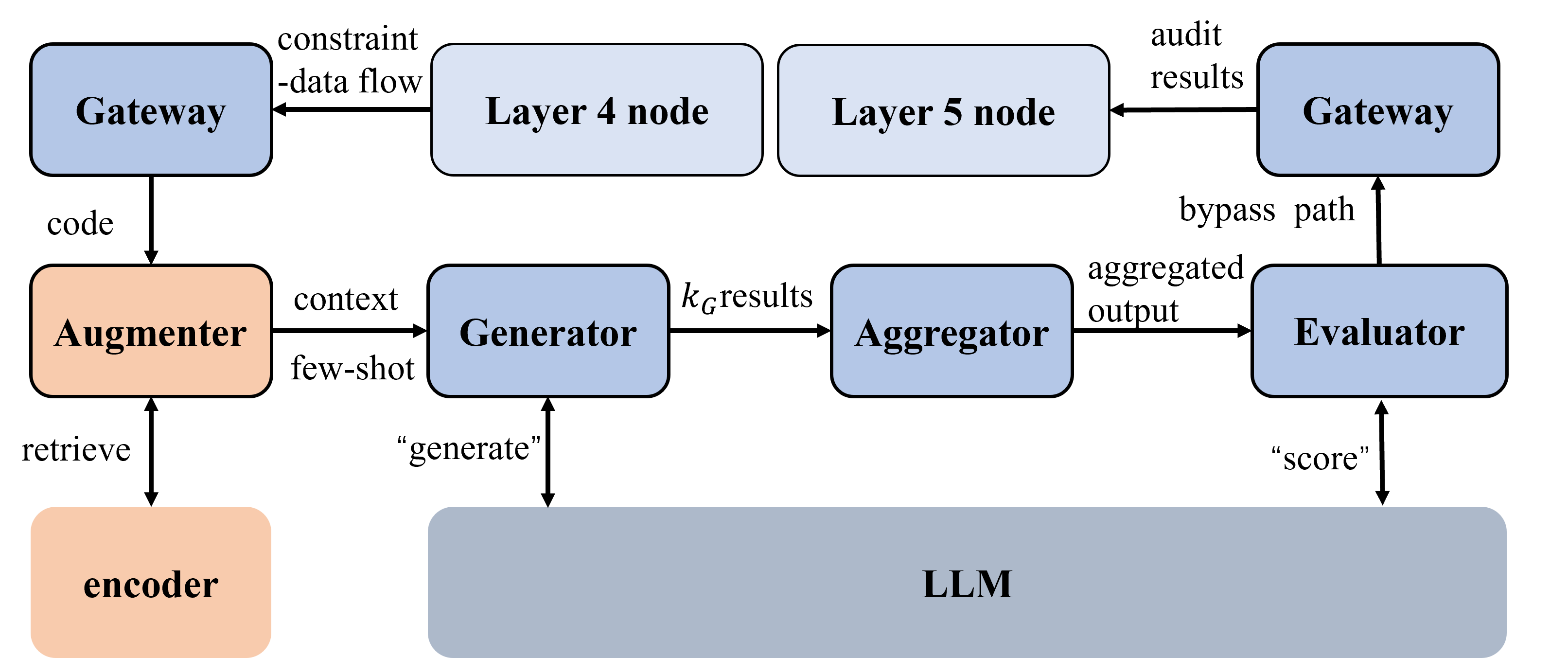}
	\caption{Workflow of Layer 5}
\end{figure}

\begin{figure*}[hbt]
	\centering
	\includegraphics[width=1\textwidth]{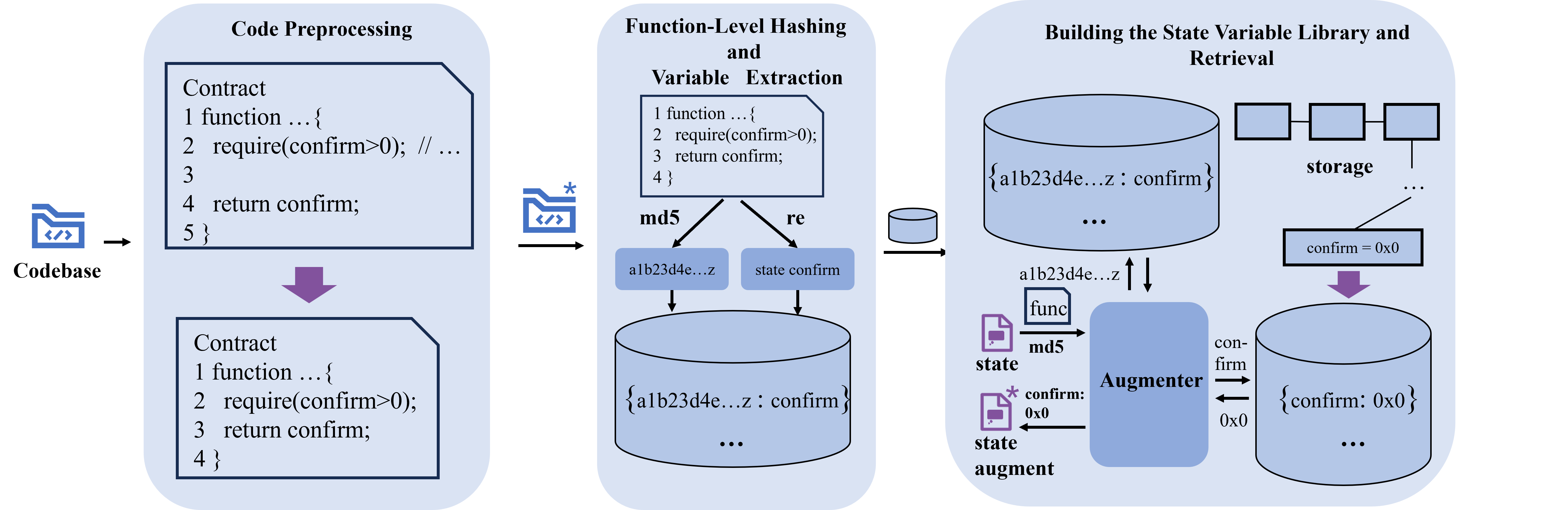}
	\caption{State Context Enhancement Method}
\end{figure*}

\subsection{Implementation}
The GoAT-X architecture is implemented in roughly 7,000 lines of Python, with the following key specifications.

\noindent \textbf{Hallucination Mitigation}.
To mitigate hallucinations in LLM-generated outputs, we go beyond ensemble-based aggregation and structured prompting by incorporating an explicit self-correction mechanism. We introduce a lightweight error detection module that validates each output against the Gateway-defined verification framework. Outputs are deemed invalid if they violate the JSON schema, omit required fields, or introduce content unsupported by the input. Detected errors are incorporated into a self-correction prompt, which is then appended to the original prompt and fed back to the model for iterative refinement. Models that fail to produce a valid output within a bounded number of attempts are discarded to prevent unreliable analyses from propagating downstream.
\noindent
\begin{tcolorbox}[enhanced, colback=gray!15, colframe=black, sharp corners, boxrule=0.6pt, left=2pt, right=2pt, top=1.5pt, bottom=2pt]
	\textbf{Self-Correction Prompt} \\ \relax
	[System Error]: The previous generation failed with error: \{error\_msg\}. Please strictly follow the format requirements, ensure the output is valid JSON, and the required fields contain valid and correct information.
\end{tcolorbox}
\noindent \textbf{Model and its parameter}. For our experiments, we select a representative ensemble of mainstream closed-source and open-source LLMs, including DeepSeek \cite{DeepSeekChat2025}, GPT \cite{OpenAIChatGPT2025}, and Gemini \cite{GoogleGemini2025}. All selected models support context windows of at least 64k tokens, satisfying our experimental requirements. To balance performance, cost, and efficiency, we adopt stable and cost-effective variants from each family: DeepSeek-V3.2, Gemini-2.5-Flash, and GPT-4o-Mini. We exclude DeepSeek-Reasoner due to its excessive deliberation behavior, which results in slower inference and increased hallucination rates \cite{si2025excessive}, and therefore favor the more stable DeepSeek-V3.2. In addition, based on its superior empirical performance in our experiments, we select Gemini as the Evaluator model throughout the system.

For the experimental configuration, we set a uniform temperature of 0.7 for all models, while all other parameters were kept at their respective official default settings, as top-p is 1, frequency penalty and presence penalty are both 0.

\noindent \textbf{Static analysis tool}. In the static analysis module, we employ two tools: Slither \cite{feist2019slither} and Surya \cite{ConsenSysDiligence_Surya_2025}. Slither, a well-established smart contract analyzer \cite{ghaleb2020effective}, serves as our primary tool due to its precise semantic analysis. However, Slither requires successful compilation, which is often infeasible for cross-chain contract codebases lacking complete dependencies. To overcome this limitation, we incorporate Surya, which extracts abstract syntax tree (AST) structures and function call relationships without requiring compilation. By combining these tools, we reliably recover ASTs and call graphs for complex cross-chain contract workflows.

\noindent \textbf{Experimental Setup}.  We set the number of generator models to $k_\mathcal{G}=3$. A global confidence threshold of 60 is applied to aggressively prune low-confidence analysis paths. For semantic pruning, we use a cosine similarity threshold of 0.85 to eliminate highly redundant candidates, ensuring pruning precision. In contrast, the similarity threshold for Bypass Knowledge Retrieval is set to 0.5, favoring broader semantic coverage so as to provide sufficiently diverse and relevant bypass rationales to the LLM. This configuration balances computational efficiency with analysis accuracy. All experiments were conducted on a server equipped with an Intel Xeon Gold 6146 CPU, an NVIDIA Tesla V100S PCIe GPU, and 64 GB of system memory.

\section{Evaluation}
\label{sec:evaluation}
\noindent \textbf{Benchmark}. To construct a comprehensive and up-to-date benchmark of vulnerabilities found in cross-chain codebases, we began by integrating findings from existing academic vulnerability research \cite{liao2024smartaxe,augusto2024xchainwatcher,zhang2024attack}. We then expanded this dataset by collecting and curating public incident reports from authoritative security audit firms and platforms like Chainlink\cite{Chainlink2025}, CertiK\cite{CertiK2022CrossChainVulnerabilities}, and Immunefi\cite{Immunefi2023CommonCrossChainBridgeVulns}. The resulting dataset provides a comprehensive benchmark of vulnerabilities specific to cross-chain token transactions. It includes real-world cases from 20 distinct cross-chain bridge projects, covering a total of \textbf{673 smart contracts}, as detailed in Table \ref{tab:vuln_results_optimized}. Each incident is meticulously annotated with its problem description and root cause, establishing a rigorous ground truth for evaluation.

In this work, we aim to answer the following research questions (RQs):

\begin{itemize}
	\item \textbf{RQ1}: Can GoAT-X accurately capture the semantic logic of cross-chain interactions to derive correct constraints and bypasses?
	\item \textbf{RQ2}: How effective is GoAT-X in detecting vulnerabilities on benchmark datasets? Can GoAT-X achieve state-of-the-art results?
	\item \textbf{RQ3}: What is the runtime efficiency and financial costs of GoAT-X?
	\item \textbf{RQ4}: How accurately can GoAT-X uncover real-world business logic risks in cross-chain bridge projects?
\end{itemize}
\subsection{RQ1:Semantic Accuracy of GoAT-X Audit Point Identification}

To answer RQ1,we established a rigorous ground truth through expert manual annotation across 20 cross-chain projects comprising 673 smart contracts, resulting in 294 distinct fine-grained audit points. As detailed in our Verification Framework, annotations were categorized into three semantic layers: (1) Property-Parameter Mapping (mapping properties to code variables), (2) Predicate Coverage Check (aligning predicates with implementation logic), and (3) Bypass Analysis (identifying vulnerabilities and rationales).

Performance was measured via Precision, Recall, and F1-score. We recorded a True Positive (TP) only when the model's output semantically aligned with the expert annotation, including the correct prediction of non-existent links (`None'). A False Negative (FN) was logged for any missed annotation, and a False Positive (FP) for any reported finding not present in the ground truth. Then we calculate the corresponding recall = $TP/(TP+FN)$, precision = $TP/(TP + FP)$, F1-score = $2*recall*precision/(recall + precision)$. 

To demonstrate the effectiveness of our design, we conduct a comprehensive ablation study to systematically assess the contribution of each major component. In addition to the full GoAT-X framework with model ensembling, we evaluate: (1) single-model baselines using DeepSeek-V3.2, GPT-4o-Mini, and Gemini-2.5-Flash, where model ensembling is disabled but the graph-based reasoning framework is retained and instantiated via iterative queries to a single model; (2) a chain-of-thought variant, CoAT-X, which replaces the graph-based reasoning in GoAT-X with a linear chain structure and is instantiated with Gemini-2.5-Flash; and (3) a GoAT-X variant with the RAG modules in Layer 5 removed. The results of these ablation experiments are reported in Table \ref{audinting performance}.

\begin{table}[htbp]
	\centering
	\setlength{\tabcolsep}{3.5pt}
	\caption{Auditing Performance with Different Methods}
	\label{audinting performance}
	\begin{tabular}{cccc|ccc}
		\toprule
		\small
		\textbf{Method} &  \textbf{\#TP}  & \textbf{\#FN} & \textbf{\#FP} & \textbf{Recall}  & \textbf{Precision} & \textbf{F1-score}\\
		\midrule
		GoAT-X & 270 & 24 & 25& \textbf{0.92} & 0.92 & \textbf{0.92} \\
		GoAT-X(DeepSeek) & 244 & 50 & 9 &0.83  & \textbf{0.96}  &  0.89 \\
		GoAT-X(GPT) & 251 & 43 & 68 & 0.85 & 0.79 & 0.82 \\
		GoAT-X(Gemini) & 257 & 37 & 18 & 0.87 & 0.93 & 0.90 \\
		CoAT-X(Gemini) & 247 & 47 & 15 & 0.84 & 0.94 & 0.89 \\
		GoAT-X w/o RAG & 261 & 33 & 22 & 0.89 & 0.92 & 0.90 \\

		\bottomrule
	\end{tabular}
\end{table}

The experimental results show that GoAT-X accurately identifies the vast majority of annotated vulnerability points, achieving the best trade-off between precision and coverage and thereby validating the effectiveness of the overall architecture.

While single-model variants remain effective under graph-based reasoning, they exhibit clear recall–precision biases due to model-specific behaviors. Model ensembling mitigates these systematic biases and significantly improves overall robustness.

Moreover, graph-based reasoning consistently outperforms linear chain reasoning in terms of recall, particularly for complex bypass logic, while maintaining comparable precision. Finally, the RAG module provides critical high-level bypass cues that are difficult to infer from code structure alone, enabling the system to recover key vulnerability points missed in the final analysis layer.

\begin{tcolorbox}[
	colback=black!5,  
	colframe=black!20,
	boxrule=1pt,      
	arc=8pt,          
	boxsep=5pt,       
	]
	\textbf{Answer to RQ1:} GoAT-X demonstrates strong semantic understanding in accurately identifying fine-grained audit points. Leveraging model ensembling, it achieves 92\% coverage of expert-annotated ground-truth vulnerability points. Extensive ablation studies further validate the effectiveness and necessity of each major component in the overall architecture.
\end{tcolorbox}

\subsection{RQ2:Effectiveness of Vulnerability Detection}
To answer RQ2, we evaluate GoAT-X's ability to identify the root causes of vulnerabilities within an entire project. Since our methodology does not detect vulnerabilities directly, the criteria for defining TP, FN, and FP are adapted accordingly. A TP is recorded if the missing rules and bypass paths reported by GoAT-X, once remediated, would prevent relevant vulnerabilities in that project. An FP is recorded if GoAT-X reports non-existent issues (e.g., proposes an incorrect missing rule or an invalid bypass path). An FN is recorded if GoAT-X fails to identify the critical missing rules or bypasses that are the root cause of the known vulnerabilities. All generated reports were then manually audited by our security experts to determine the final results.

As delineated in Table \ref{tab:baselines_comparison}, our evaluation involves a comparative analysis with several state-of-the-art tools. We strictly control for experimental variables by configuring GPTScan and SmartInv with the same underlying model (Gemini-2.5-Flash) and input nodes as GoAT-X. For SmartAxe, whose open-source implementation is incomplete, we adopt a literature-based comparison; this approach is methodologically sound as our evaluation benchmark encompasses the projects analyzed in their original work, ensuring a valid and direct comparison of detection performance. Notably, PropertyGPT is excluded due to its lack of cross-chain semantics and specialized training datasets.

To further assess the necessity of GoAT-X’s layered design and iterative reasoning process, we introduce two simplified ablation baselines: (1) \textbf{Direct-Audit}, where the LLM performs a zero-shot cross-chain audit on the extracted code nodes, and (2) \textbf{Single-Step Analysis}, where the LLM attempts to synthesize parameter mapping, rule checking, and bypass identification in a single pass without iterative refinement.

\begin{table}[t]
	\centering
	\caption{Comparison of Baselines and Experimental Configurations.}
	\label{tab:baselines_comparison}
	\setlength{\tabcolsep}{4pt} 
	\footnotesize 
	\begin{tabular}{l|c|c|c|l}
		\toprule
		\textbf{Tool} & \textbf{Type} & \textbf{X-chain\textsuperscript{*}} & \textbf{Avail.} & \textbf{Comparison Strategy} \\ 
		\midrule
		\textbf{GoAT-X} & Hybrid & $\checkmark$ & $\checkmark$ & Main Evaluation \\ 
		\midrule
		GPTScan~\cite{sun2024gptscan} & Hybrid & $\times$ & $\checkmark$ & Direct Execution\textsuperscript{\dag} \\ 
		SmartInv~\cite{wang2024smartinv} & LLM & $\times$ & $\checkmark$ & Direct Execution\textsuperscript{\dag} \\ 
		SmartAxe~\cite{liao2024smartaxe} & Static & $\checkmark$ & $\times$ & Literature-based \\ 
		PropertyGPT~\cite{liu2024propertygpt} & Hybrid & $\times$ & $\times$ & Excluded (Incompatible) \\ 
		\bottomrule
	\end{tabular}
	\begin{flushleft}
		\scriptsize \textsuperscript{*}X-chain: Cross-chain Focus; \textsuperscript{\dag}Executed using Gemini-2.5-Flash.
	\end{flushleft}
\end{table}

\begin{table*}[t]
	\centering
	\setlength{\tabcolsep}{3pt}
	
	\small
	
	\caption{The table shows the identification results for each cross-chain project with known vulnerabilities, where the symbols represent the framework's overall performance: \CIRCLE indicates perfect detection (TP); \LEFTcircle   indicates an Over-report(TP + FP); \RIGHTcircle indicates an Under-report(FN); and \Circle indicates a complete failure(FN + FP).}
	\label{tab:vuln_results_optimized}
	
	\begin{tabular*}{\textwidth}{@{\extracolsep{\fill}} l *{20}{c}}
		\toprule
		
		\textbf{Method} & 
		\rot{Allbridge} & 
		\rot{ChainSwap} & 
		\rot{HyperBridge} & 
		\rot{LIFI I} & 
		\rot{LIFI II} & 
		\rot{MeterPassPort} & 
		\rot{Multichain I} & 
		\rot{Multichain II} & 
		\rot{Nerve} & 
		\rot{Nomad} & 
		\rot{pNetwork} & 
		\rot{PolyNetwork} & 
		\rot{QBridge} & 
		\rot{Qubit} & 
		\rot{Ronin} & 
		\rot{Rubic} & 
		\rot{SocketGateway} &  
		\rot{Synapse}  &
		\rot{THORChain} & 
		\rot{XBridge} \\
		
		\midrule
		

		Smartlnv(Gemini) & \Circle & \Circle & \Circle & \LEFTcircle & \LEFTcircle & \Circle & \CIRCLE & \Circle & \Circle & \Circle & \Circle & \Circle & \Circle & \Circle & \Circle & \CIRCLE & \CIRCLE & \Circle & \Circle & \Circle \\
		
		GPTScan(Gemini) & \RIGHTcircle & \RIGHTcircle & \RIGHTcircle & \CIRCLE & \RIGHTcircle & \RIGHTcircle & \RIGHTcircle & \RIGHTcircle & \CIRCLE & \RIGHTcircle & \RIGHTcircle & \RIGHTcircle & \RIGHTcircle & \RIGHTcircle & \RIGHTcircle & \RIGHTcircle & \RIGHTcircle & \CIRCLE & \RIGHTcircle & \RIGHTcircle \\
		
		Direct-Audit & \Circle & \RIGHTcircle & \RIGHTcircle & \CIRCLE & \LEFTcircle & \Circle & \RIGHTcircle & \RIGHTcircle & \Circle & \Circle & \RIGHTcircle & \Circle & \LEFTcircle & \RIGHTcircle & \RIGHTcircle & \CIRCLE & \LEFTcircle & \Circle & \RIGHTcircle & \Circle \\
		
		Single-Step Analysis & \Circle & \Circle & \RIGHTcircle & \Circle & \CIRCLE & \RIGHTcircle & \Circle & \RIGHTcircle & \Circle & \RIGHTcircle & \CIRCLE & \CIRCLE & \RIGHTcircle & \RIGHTcircle & \RIGHTcircle & \RIGHTcircle & \LEFTcircle & \LEFTcircle & \CIRCLE & \RIGHTcircle \\
		
		GoAT-X & \RIGHTcircle & \CIRCLE & \CIRCLE & \CIRCLE & \CIRCLE & \LEFTcircle & \CIRCLE & \CIRCLE & \LEFTcircle & \CIRCLE & \CIRCLE & \CIRCLE & \CIRCLE & \CIRCLE & \LEFTcircle & \CIRCLE & \CIRCLE & \LEFTcircle & \CIRCLE & \CIRCLE   \\
		
		\bottomrule
	\end{tabular*}
\end{table*}

\begin{table}[htbp]
	\centering
	\caption{Performance Comparison of Methods}
	\label{tab:vuln_results_optimized_metric}
	\begin{tabular}{cccc}
		\toprule
		\textbf{Method} & \textbf{Recall}  & \textbf{Precision} & \textbf{F1-score}\\
		\midrule
		SmartInv(Gemini) & 0.25  & 0.23  &  0.24 \\
		GPTScan(Gemini) & 0.15  & \textbf{1.00}  &  0.26 \\
		SmartAxe & 0.90 & 0.85 & 0.87 \\
		Direct-Audit & 0.25 & 0.33 & 0.29 \\
		Single-Step Analysis & 0.30 & 0.46 & 0.36 \\
		GoAT-X & \textbf{0.95}  & 0.83  &  0.88 \\
		
		\bottomrule
	\end{tabular}
	
\end{table}

The comparative results are summarized in Table \ref{tab:vuln_results_optimized} and Table 
\ref{tab:vuln_results_optimized_metric}. Overall, the evaluation demonstrates the limitations of existing tools when applied to cross-chain vulnerability analysis. As shown in Table~\MakeUppercase{\romannumeral 5}, GoAT-X significantly outperforms all baselines, achieving a Recall of 0.95, a Precision of 0.83, and an F1-score of 0.88. In contrast, the LLM-based auditors SmartInv and GPTScan perform poorly, with F1-scores of 0.24 and 0.26, respectively. GPTScan’s perfect Precision is a byproduct of its extremely low Recall (0.15), indicating that most critical issues remain undetected.

While SmartAxe reports competitive results in isolated settings, it still underperforms GoAT-X, reflecting the inherent difficulty static analysis faces in resolving cross-chain business logic and state dependencies. Moreover, unlike low-level alerts produced by static analyzers, GoAT-X generates structured natural-language reports that improve the interpretability and actionability of audit results.

The baselines further highlight the necessity of GoAT-X’s hierarchical decomposition and iterative reasoning strategy. Even when provided with the same extracted cross-chain logic, both \textit{Direct-Audit} and \textit{Single-Step Analysis} fail to achieve reliable detection. This confirms that neither coarse-grained code extraction nor single-pass reasoning is sufficient; precise data-flow slicing and decomposed reasoning are essential for effective cross-chain vulnerability detection.

To understand the boundaries of our framework, we analyzed the single False Negative case observed in the \textit{Allbridge} project. The vulnerability in \textit{Allbridge} stemmed from a flash loan attack caused by insufficient slippage protection. Our analysis revealed that the project did not use explicit state variables for slippage checks; instead, it relied on implicit calculations via auxiliary functions (e.g., \texttt{getY}). This implicit data flow obscured the semantic link between the calculation logic and the concept of ``slippage", preventing the model from inferring the necessary constraint. This case highlights a persistent challenge in smart contract auditing: resolving implicit semantic dependencies. Future work will address this by enhancing the framework's ability to track semantic data flows or through targeted fine-tuning on implicit arithmetic patterns.

\begin{tcolorbox}[
	colback=black!5,  
	colframe=black!20,
	boxrule=1pt,      
	arc=8pt,          
	boxsep=5pt,       
	]
	\textbf{Answer to RQ2:} GoAT-X is highly effective, achieving a peak F1-score of 88\% and detecting nearly all vulnerabilities with a 95\% Recall. Comparative analysis confirms its state-of-the-art performance, as it not only significantly outperforms existing LLM-based auditors but also surpasses the specialized static analysis tool.
\end{tcolorbox}

\subsection{RQ3:Analysis of Runtime Efficiency and Financial Cost}
In RQ3, to evaluate the efficiency and cost-effectiveness of GoAT-X, we conducted detailed measurements for the detection process of each project. We recorded the runtime of each layer within the framework, as well as the number of tokens consumed and the corresponding financial cost incurred by the LLM Layers. The number of tokens per project is shown in Figure 10, the layer-by-layer time consumption is presented in Figure 11, and the final average results are summarized in \ref{Measurement}.

\begin{figure}[H]
	\centering
	\includegraphics[width=0.5\textwidth]{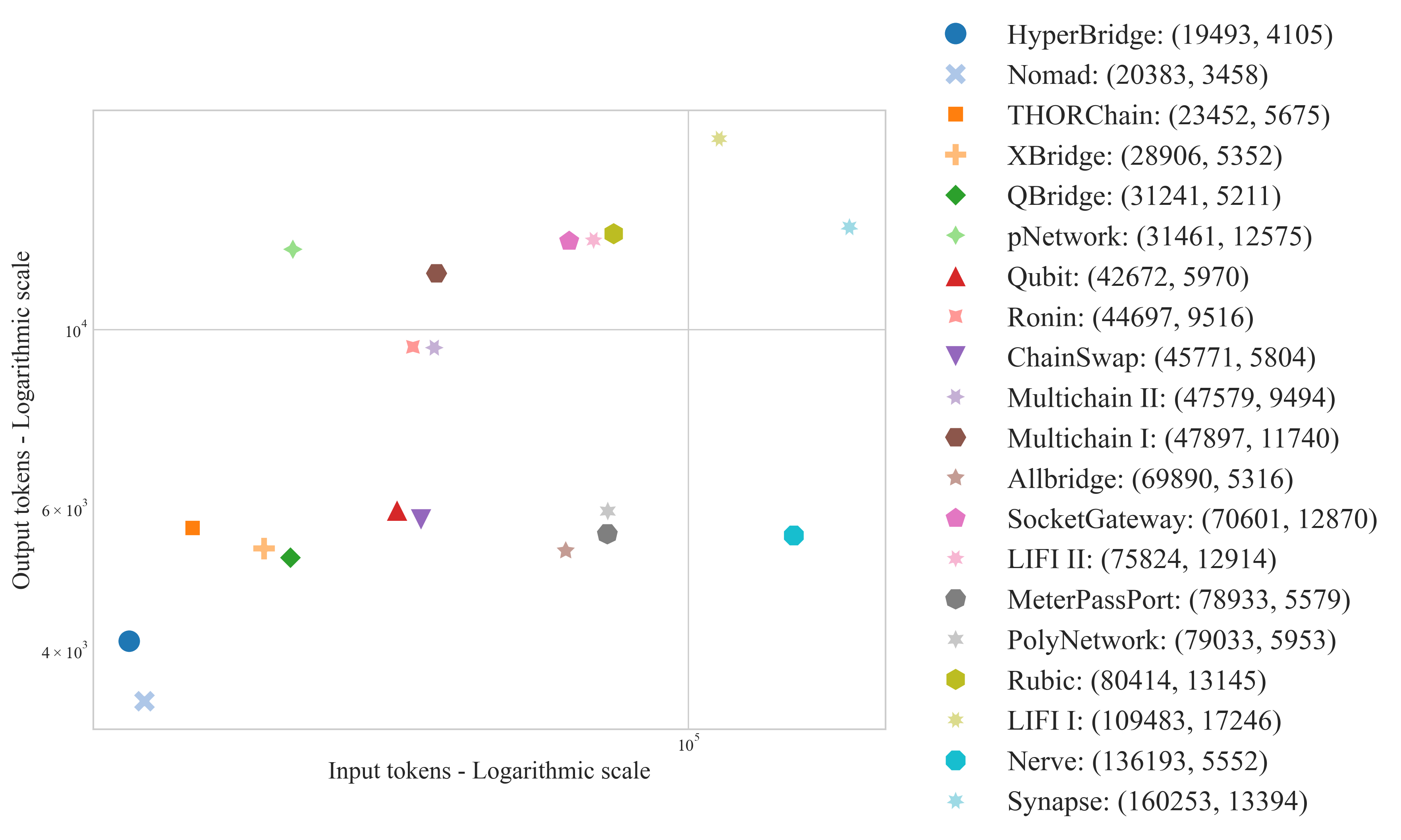}
	\caption{Input and output token counts for each project, plotted on a logarithmic scale. In the legend, (x, y) denotes the number of input tokens (x) and output tokens (y), respectively.}
\end{figure}

\begin{figure}[H]
	\centering
	\includegraphics[width=0.5\textwidth]{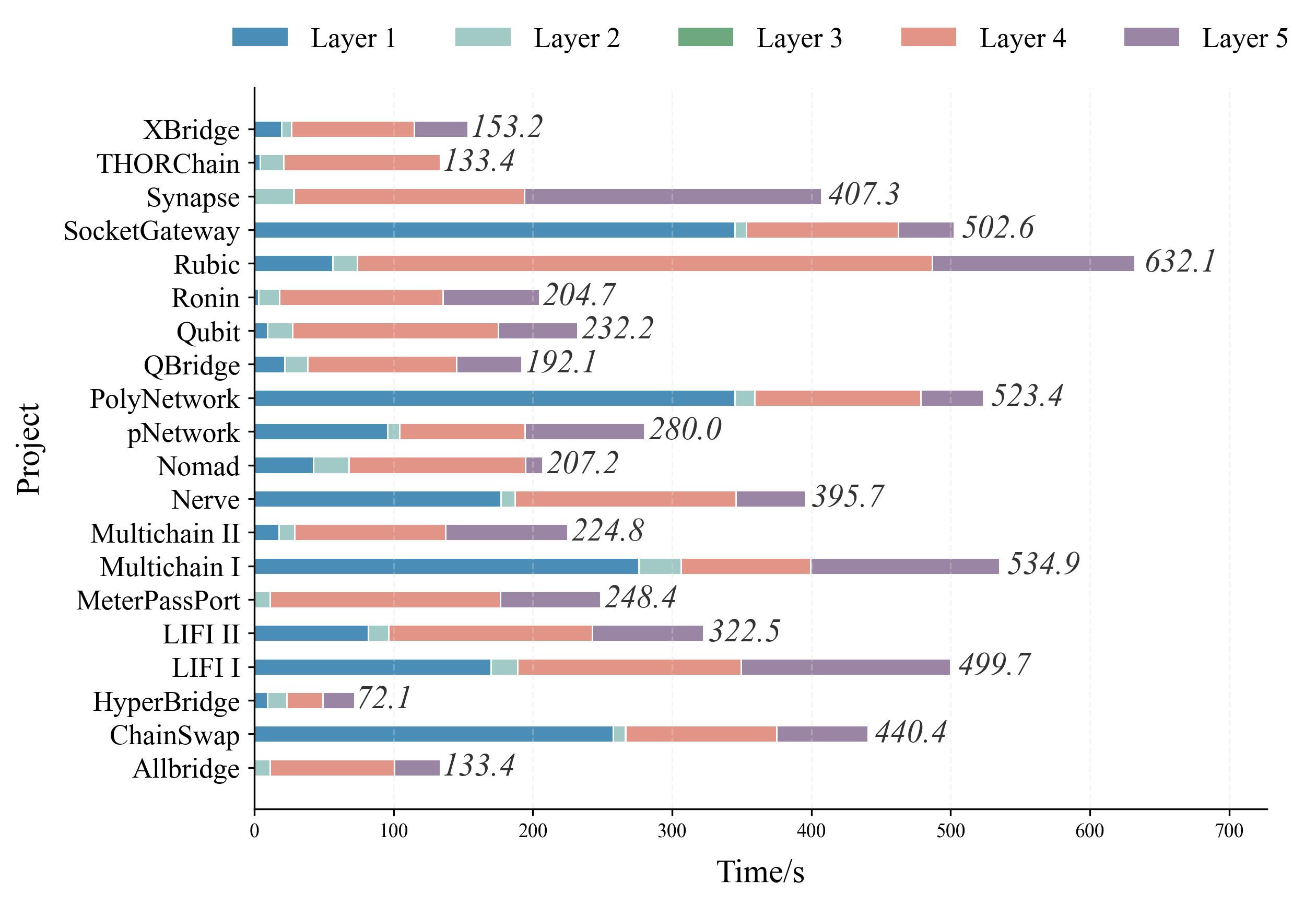}
	\caption{Runtime of Each Layer in GoAT-X}
\end{figure}

\begin{table}[htbp]
	\centering
	\caption{Average Measurement Per Project}
	\label{Measurement}
	
	\begin{tabular}{ccccc}
		\toprule
		\textbf{Method} & \textbf{Time}  & \textbf{Input tokens} & \textbf{Output tokens} &\textbf{Costs} \\
		\midrule
		GoAT-X & 317.02 s & 62,208 & 8,543 & \$0.02 \\
		
		\bottomrule
	\end{tabular}
\end{table}

These results demonstrate the strong cost-effectiveness of our framework when analyzing large and complex cross-chain codebases. By leveraging the static analysis layer to precisely extract core cross-chain logic and parameter-specific data flow slices, the framework substantially reduces the token consumption required by large language models, thereby lowering both financial cost and computational overhead.

In addition, the encoder-based pruning strategy eliminates, on average, 3 branches per project that would otherwise require validation by high-cost models, further reducing unnecessary LLM invocations. With an average detection time of approximately 317 seconds per project, the framework achieves high runtime efficiency, making it well-suited for deep, automated auditing of complex cross-chain systems.
\begin{tcolorbox}[
	colback=black!5,  
	colframe=black!20,
	boxrule=1pt,      
	arc=8pt,          
	boxsep=5pt,       
	]
	\textbf{Answer to RQ3:}  At an average auditing overhead of just \$0.02 per project, GoAT-X proves to be highly cost-effective. Concurrently, its average detection time of approximately 317 seconds per project validates its efficiency, making it a viable tool for offline security analysis.
\end{tcolorbox}

\subsection{RQ4:Impact of GoAT-X}
In RQ4, to evaluate GoAT-X's efficacy in realistic scenarios, we constructed a cross-chain wild dataset comprising \textbf{1,406} smart contracts from 12 open-source projects. We applied GoAT-X to scan these projects and manually verified the reported alerts. As summarized in Table \ref{wilddataset}, the tool successfully identified critical security hotspots where constraints were either missing or potentially bypassable, demonstrating its capability to uncover deep business logic risks in complex environments.

\begin{table}[hbt]
	\centering
	\caption{Detection Results on the Cross-Chain Wild Dataset}
	\label{wilddataset}
	
	\begin{tabular}{l|ccc|c}
		\toprule
		\small
		\textbf{Category} & \textbf{\#Alerts} & \textbf{\#Confirmed} & \textbf{\#FP} & \textbf{Precision} \\
		\midrule
		Missing Constraints & 116 & 108 & 8 & 0.93 \\
		Constraint Bypass & 12 & 9 & 3 & 0.75 \\
		\midrule
		\textbf{Total} & \textbf{128} & \textbf{117} & \textbf{11} & \textbf{0.91} \\
		\bottomrule
	\end{tabular}
\end{table}

We acknowledge that not all flagged risks correspond to immediately exploitable vulnerabilities. In practice, some identified issues represent deviations from strict defensive programming, such as relying on implicit compiler-level checks (e.g., Solidity 0.8.0+ for overflow) rather than explicit validation. However, we argue that maintaining security redundancy  is crucial for cross-chain protocols, where minor oversights can lead to catastrophic losses. GoAT-X operates as a strict auditor, flagging these potential risks to ensure comprehensive coverage. Below, we present two representative cases from our findings (anonymized for security):

\noindent \textbf{Case 1:} As shown in Figure 12, the transition from $withdraw$ to the internal $\_withdrawTo$ function validates the $secret$ but fails to check if the nonce has been consumed. This Missing Constraint exposes the protocol to Replay Attacks, where an attacker could potentially reuse a valid $secret$ to drain funds repeatedly.

\begin{figure}[H]
	\centering
	\includegraphics[width=0.42\textwidth]{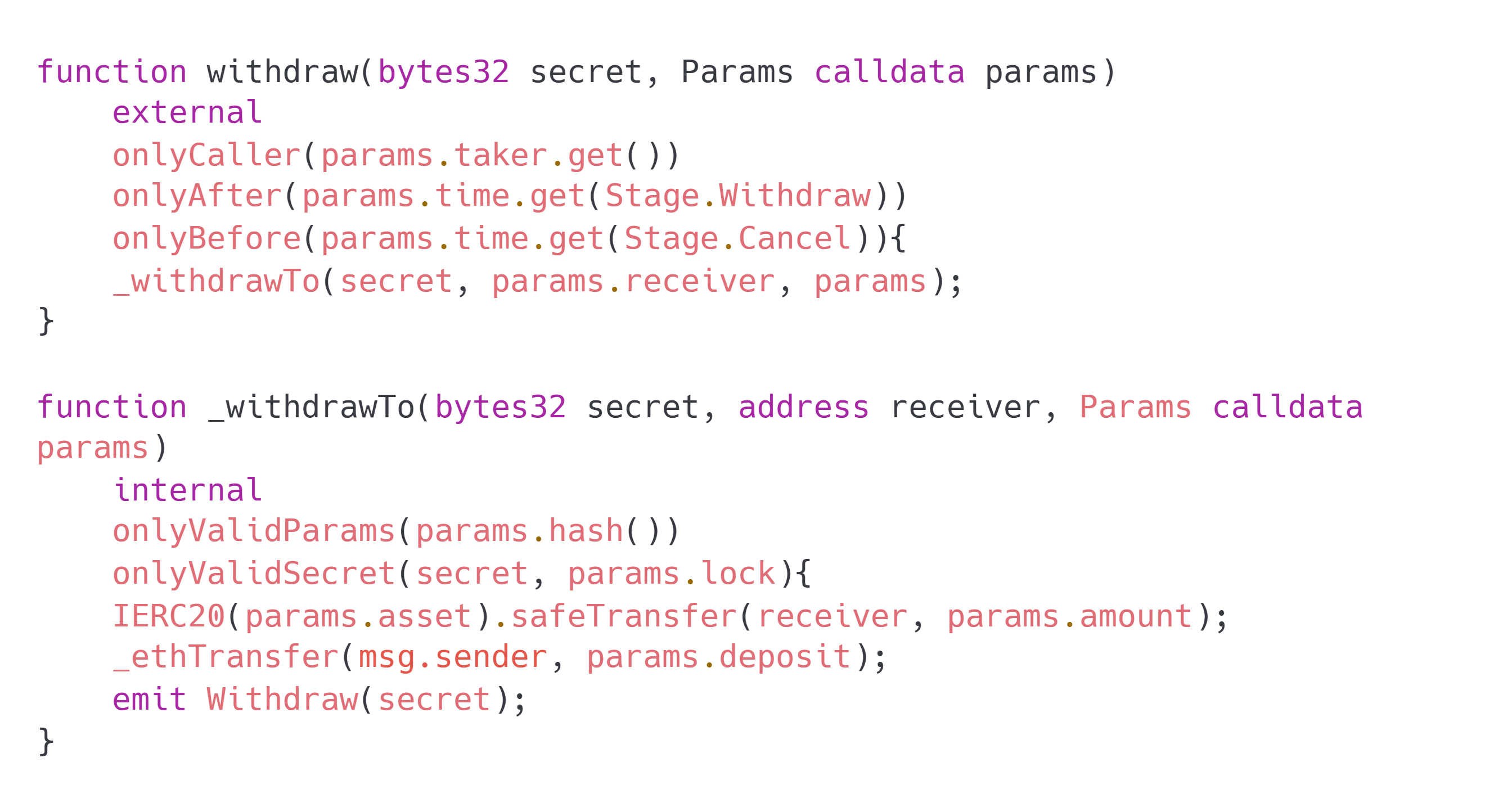}
	\caption{Instance of Missing Constraint}
\end{figure}

\noindent \textbf{Case 2:} As illustrated in Figure 13, GoAT-X identifies a high-risk data flow in which user-supplied $callData$ is forwarded into a low-level call. Although the contract enforces a whitelist on the function selector (i.e., the first four bytes of the $callData$), the remaining arguments remain fully user-controlled. This design introduces a fragile reliance on downstream validation: if subsequent checks are incomplete or improperly implemented, an attacker may manipulate these arguments to craft malformed external calls. GoAT-X correctly flags this execution path as a critical validation bypass, as such manipulation can directly enable unauthorized token transfers and result in severe economic losses.

\begin{figure}[H]
	\centering
	\includegraphics[width=0.42\textwidth]{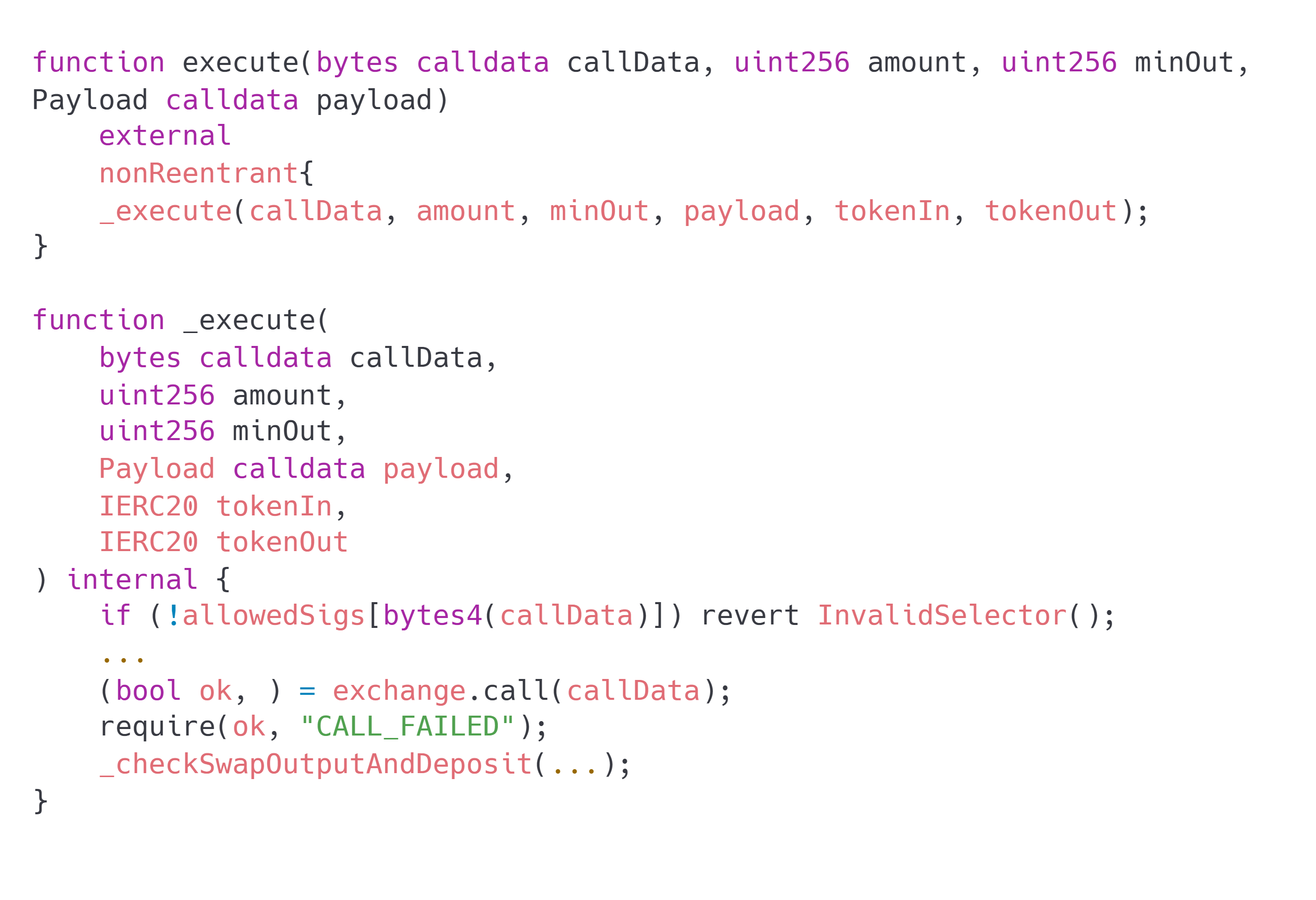}
	\caption{Instance of Constraint Bypass}
\end{figure}

\begin{tcolorbox}[
	colback=black!5,  
	colframe=black!20,
	boxrule=1pt,      
	arc=8pt,          
	boxsep=5pt,       
	]
	\textbf{Answer to RQ4:}  GoAT-X demonstrated high precision in the wild dataset. Out of the total alerts, we confirmed 108 instances of missing constraints and 9 valid bypass paths, proving its practical value in identifying realistic threats.
\end{tcolorbox}

\section{Related work}
\noindent \textbf{Cross-Chain Vulnerability Detection.} Current research on securing cross-chain bridges primarily follows two main approaches. The first is pre-deployment static analysis, which operates at the bytecode level but suffers from a semantic gap, where high-level business logic is obscured, making it difficult to discern the intent behind key operations \cite{liao2024smartaxe,wang2024xguard}. The second is post-deployment monitoring, which detects attacks in real-time but is inherently reactive rather than preventative due to the immutability of smart contracts\cite{zhang2022xscope,augusto2024xchainwatcher,belchior2023hephaestus}.  

GoAT-X transcends these limitations by operating directly at the source-code level. It synergizes the precision of static analysis with the semantic understanding of Large Language Models, all of which is orchestrated by our formal verification framework, to generate a detailed audit report that captures high-level business logic.

\noindent \textbf{Single-Chain Tools.} Outside of cross-chain contexts, a variety of automated tools have been developed to uncover vulnerabilities in smart contracts. Symbolic execution frameworks, such as Oyente \cite{luu2016making}, Osiris \cite{torres2018osiris}, Mythril \cite{mueller2017framework}, and Manticore \cite{mossberg2019manticore}, explore all feasible program paths by substituting concrete values with symbolic variables. Static analyzers like Slither \cite{feist2019slither} and Surya \cite{ConsenSysDiligence_Surya_2025} traverse the contract’s abstract syntax tree to identify patterns indicative of bugs, while fuzzers such as sFuzz \cite{nguyen2020sfuzz} and ContractFuzzer \cite{jiang2018contractfuzzer} probe runtime behaviors and execution traces for anomalous states.

Although these techniques do not directly address the challenges of cross-chain interactions, their core ideas are highly relevant. In our method, we leverage the high efficiency and precision of static analysis tools in our foundational layers to construct detailed function call graphs and extract parameter-specific data flows, which provide the essential grounded context for our deeper analysis.

\noindent \textbf{LLM-Driven Security Analysis for Smart Contract.} Current research in applying LLMs to this field spans several directions, ranging from directly querying the LLM for rapid code auditing \cite{wei2024llm, li2025scalm}, to enhancing static analysis by using the LLM to translate complex contract logic into a more analyzable intermediate representation \cite{sun2024gptscan, wu2025detecting}, and employs fine-tuning and in-context learning to automatically generate and refine formal properties for new smart contracts. \cite{liu2024propertygpt,wang2024smartinv}.

Distinguishing itself from prior art focused on single-chain analysis, GoAT-X is the first framework to leverage LLM for the automated auditing of cross-chain transaction contracts, with a design that is readily extensible to other code security domains.

\section{Conclusion}
This work addresses a critical security gap in the multi-chain ecosystem by introducing GoAT-X, a framework that brings first-principles verification to cross-chain smart contract codebases. Our study shows that effective automation of complex audits depends not on scaling LLMs alone, but on structuring and constraining their reasoning. GoAT-X achieves this by grounding generative analysis in rigorous static slicing and a formal verification framework, while leveraging ensemble reasoning and retrieval-augmented guidance to stabilize inference and reduce spurious conclusions. Together, these designs transform vulnerability discovery from an ad hoc, stochastic process into a systematic and reproducible verification workflow. Our evaluation confirms GoAT-X’s superiority, achieving 95\% coverage of vulnerable projects and uncovering 117 confirmed risks in the wild with exceptional cost-efficiency. This work establishes a scalable auditing paradigm for blockchain security, with promising applicability to broader code analysis domains and the sustainable development of the DeFi ecosystem.

\section*{Acknowledgments}
This work was supported in part by the National Natural Science Foundation of China (NSFC) / Research Grants Council (RGC) Collaborative Research Scheme (Grant No. 62461160332 \& CRS\_HKUST602/24), and in part by the Major Key Project of Peng Cheng Laboratory under Grant PCL2025A07.


\bibliographystyle{IEEEtran}
\bibliography{references}

\begin{thebibliography}{10}
\providecommand{\url}[1]{#1}
\csname url@samestyle\endcsname
\providecommand{\newblock}{\relax}
\providecommand{\bibinfo}[2]{#2}
\providecommand{\BIBentrySTDinterwordspacing}{\spaceskip=0pt\relax}
\providecommand{\BIBentryALTinterwordstretchfactor}{4}
\providecommand{\BIBentryALTinterwordspacing}{\spaceskip=\fontdimen2\font plus
\BIBentryALTinterwordstretchfactor\fontdimen3\font minus
  \fontdimen4\font\relax}
\providecommand{\BIBforeignlanguage}[2]{{%
\expandafter\ifx\csname l@#1\endcsname\relax
\typeout{** WARNING: IEEEtran.bst: No hyphenation pattern has been}%
\typeout{** loaded for the language `#1'. Using the pattern for}%
\typeout{** the default language instead.}%
\else
\language=\csname l@#1\endcsname
\fi
#2}}
\providecommand{\BIBdecl}{\relax}
\BIBdecl

\bibitem{ante2021influence}
L.~Ante, I.~Fiedler, and E.~Strehle, ``{The Influence of Stablecoin Issuances
  on Cryptocurrency Markets},'' \emph{Finance Research Letters}, vol.~41, p.
  101867, 2021.

\bibitem{ou2022overview}
W.~Ou, S.~Huang, J.~Zheng, Q.~Zhang, G.~Zeng, and W.~Han, ``{An Overview on
  Cross-Chain: Mechanism, Platforms, Challenges and Advances},'' \emph{Computer
  Networks}, vol. 218, p. 109378, 2022.

\bibitem{belchior2023hephaestus}
R.~Belchior, P.~Somogyvari, J.~Pfannschmidt, A.~Vasconcelos, and M.~Correia,
  ``{Hephaestus: Modeling, Analysis, and Performance Evaluation of Cross-Chain
  Transactions},'' \emph{IEEE Transactions on Reliability}, vol.~73, no.~2, pp.
  1132--1146, 2023.

\bibitem{lee2023sok}
S.-S. Lee, A.~Murashkin, M.~Derka, and J.~Gorzny, ``{SoK: Not Quite Water Under
  the Bridge: Review of Cross-Chain Bridge Hacks},'' in \emph{2023 IEEE
  International Conference on Blockchain and Cryptocurrency (ICBC)}.\hskip 1em
  plus 0.5em minus 0.4em\relax IEEE, 2023, pp. 1--14.

\bibitem{Chainlink2025}
{Chainlink}, ``{Seven Key Cross-Chain Bridge Vulnerabilities Explained},''
  \url{https://chain.link/education-hub/cross-chain-bridge-vulnerabilities}.

\bibitem{slowmist2023blockchain}
{SlowMist}, ``{2023 Blockchain Security and Anti-Money Laundering Annual
  Report},'' SlowMist Security, Tech. Rep., 2024,
  \url{https://www.slowmist.com/report/2023-Blockchain-Security-and-AML-Annual-Report(EN).pdf},
  Accessed: 2025-07-28.

\bibitem{liao2024smartaxe}
Z.~Liao, Y.~Nan, H.~Liang, S.~Hao, J.~Zhai, J.~Wu, and Z.~Zheng, ``{SmartAxe:
  Detecting Cross-Chain Vulnerabilities in Bridge Smart Contracts via
  Fine-Grained Static Analysis},'' \emph{Proceedings of the ACM on Software
  Engineering}, vol.~1, no. FSE, pp. 249--270, 2024.

\bibitem{wang2024xguard}
K.~Wang, Y.~Li, C.~Wang, J.~Gao, Z.~Guan, and Z.~Chen, ``{XGuard: Detecting
  Inconsistency Behaviors of Cross-Chain Bridges},'' in \emph{Companion
  Proceedings of the 32nd ACM International Conference on the Foundations of
  Software Engineering}, 2024, pp. 612--616.

\bibitem{sheng2025llms}
Z.~Sheng, Z.~Chen, S.~Gu, H.~Huang, G.~Gu, and J.~Huang, ``{LLMs in Software
  Security: A Survey of Vulnerability Detection Techniques and Insights},''
  \emph{arXiv preprint arXiv:2502.07049}, 2025.

\bibitem{wei2024llm}
Z.~Wei, J.~Sun, Z.~Zhang, X.~Zhang, M.~Li, and Z.~Hou, ``{LLM-SmartAudit:
  Advanced Smart Contract Vulnerability Detection},'' \emph{arXiv preprint
  arXiv:2410.09381}, 2024.

\bibitem{li2025scalm}
Z.~Li, X.~Li, W.~Li, and X.~Wang, ``{SCALM: Detecting Bad Practices in Smart
  Contracts Through LLMs},'' in \emph{Proceedings of the AAAI Conference on
  Artificial Intelligence}, vol.~39, no.~1, 2025, pp. 470--477.

\bibitem{liu2024propertygpt}
Y.~Liu, Y.~Xue, D.~Wu, Y.~Sun, Y.~Li, M.~Shi, and Y.~Liu, ``{PropertyGPT:
  LLM-driven Formal Verification of Smart Contracts through Retrieval-Augmented
  Property Generation},'' in \emph{Proceedings of the 32nd Annual Network and
  Distributed System Security Symposium (NDSS 2025)}, San Diego, CA, USA, Feb.
  2025, distinguished Paper Award.

\bibitem{wang2024smartinv}
S.~J. Wang, K.~Pei, and J.~Yang, ``Smartinv: Multimodal learning for smart
  contract invariant inference,'' in \emph{2024 IEEE Symposium on Security and
  Privacy (SP)}.\hskip 1em plus 0.5em minus 0.4em\relax IEEE, 2024, pp.
  2217--2235.

\bibitem{OpenAIChatGPT2025}
J.~Achiam, S.~Adler, S.~Agarwal, L.~Ahmad, I.~Akkaya, F.~L. Aleman, D.~Almeida,
  J.~Altenschmidt, S.~Altman, S.~Anadkat \emph{et~al.}, ``Gpt-4 technical
  report,'' \emph{arXiv preprint arXiv:2303.08774}, 2023.

\bibitem{GoogleGemini2025}
G.~Team, R.~Anil, S.~Borgeaud, J.-B. Alayrac, J.~Yu, R.~Soricut, J.~Schalkwyk,
  A.~M. Dai, A.~Hauth, K.~Millican \emph{et~al.}, ``Gemini: a family of highly
  capable multimodal models,'' \emph{arXiv preprint arXiv:2312.11805}, 2023.

\bibitem{DeepSeekChat2025}
A.~Liu, B.~Feng, B.~Xue, B.~Wang, B.~Wu, C.~Lu, C.~Zhao, C.~Deng, C.~Zhang,
  C.~Ruan \emph{et~al.}, ``Deepseek-v3 technical report,'' \emph{arXiv preprint
  arXiv:2412.19437}, 2024.

\bibitem{yao2023tree}
S.~Yao, D.~Yu, J.~Zhao, I.~Shafran, T.~Griffiths, Y.~Cao, and K.~Narasimhan,
  ``{Tree of Thoughts: Deliberate Problem Solving with Large Language
  Models},'' \emph{Advances in Neural Information Processing Systems}, vol.~36,
  pp. 11\,809--11\,822, 2023.

\bibitem{besta2024graph}
M.~Besta, N.~Blach, A.~Kubicek, R.~Gerstenberger, M.~Podstawski, L.~Gianinazzi,
  J.~Gajda, T.~Lehmann, H.~Niewiadomski, P.~Nyczyk \emph{et~al.}, ``{Graph of
  Thoughts: Solving Elaborate Problems with Large Language Models},'' in
  \emph{Proceedings of the AAAI Conference on Artificial Intelligence},
  vol.~38, no.~16, 2024, pp. 17\,682--17\,690.

\bibitem{zhang2022xscope}
J.~Zhang, J.~Gao, Y.~Li, Z.~Chen, Z.~Guan, and Z.~Chen, ``{Xscope: Hunting for
  Cross-Chain Bridge Attacks},'' in \emph{Proceedings of the 37th IEEE/ACM
  International Conference on Automated Software Engineering}, 2022, pp. 1--4.

\bibitem{lifi_exploit_2022}
Zord4n, ``{LI.FI Smart Contract Vulnerability Post Mortem: 20th March -- The
  Exploit},'' \url{https://blog.li.fi/20th-march-the-exploit-e9e1c5c03eb9}.

\bibitem{halborn_synapse_2021}
R.~Behnke, ``{Explained: The Synapse and Nerve Bridge Hacks},''
  \url{https://www.halborn.com/blog/post/explained-the-synapse-and-nerve-bridge-hacks-november-2021}.

\bibitem{zhang2024attack}
M.~Zhang, X.~Zhang, Y.~Zhang, and Z.~Lin, ``{Security of Cross-Chain Bridges:
  Attack Surfaces, Defenses, and Open Problems},'' in \emph{Proceedings of the
  27th International Symposium on Research in Attacks, Intrusions and
  Defenses}, 2024, pp. 298--316.

\bibitem{wu2025safeguarding}
J.~Wu, K.~Lin, D.~Lin, B.~Zhang, Z.~Wu, and J.~Su, ``{Safeguarding Blockchain
  Ecosystem: Understanding and Detecting Attack Transactions on Cross-Chain
  Bridges},'' in \emph{Proceedings of the ACM on Web Conference 2025}, 2025,
  pp. 4902--4912.

\bibitem{ODAILY_QubitQBridge_2022}
Odaily, ``{Qubit Finance's QBridge Hacked for \$80 Million},''
  \url{https://www.odaily.news/post/5176008}.

\bibitem{feist2019slither}
J.~Feist, G.~Grieco, and A.~Groce, ``{Slither: A Static Analysis Framework for
  Smart Contracts},'' in \emph{2019 IEEE/ACM 2nd International Workshop on
  Emerging Trends in Software Engineering for Blockchain (WETSEB)}.\hskip 1em
  plus 0.5em minus 0.4em\relax IEEE, 2019, pp. 8--15.

\bibitem{ConsenSysDiligence_Surya_2025}
{ConsenSys Diligence}, ``{Surya: A Solidity Inspector -- Utilities for
  Exploring Solidity Smart Contracts},'' GitHub repository.

\bibitem{wang2024llmdfa}
C.~Wang, W.~Zhang, Z.~Su, X.~Xu, X.~Xie, and X.~Zhang, ``{LLMDFA: Analyzing
  Dataflow in Code with Large Language Models},'' \emph{Advances in Neural
  Information Processing Systems}, vol.~37, pp. 131\,545--131\,574, 2024.

\bibitem{xu2024faithful}
J.~Xu, H.~Fei, L.~Pan, Q.~Liu, M.-L. Lee, and W.~Hsu, ``{Faithful Logical
  Reasoning via Symbolic Chain-of-Thought},'' \emph{arXiv preprint
  arXiv:2405.18357}, 2024.

\bibitem{guo2022unixcoder}
D.~Guo, S.~Lu, N.~Duan, Y.~Wang, M.~Zhou, and J.~Yin, ``Unixcoder: Unified
  cross-modal pre-training for code representation,'' in \emph{Proceedings of
  the 60th Annual Meeting of the Association for Computational Linguistics
  (Volume 1: Long Papers)}, 2022, pp. 7212--7225.

\bibitem{swc_registry}
``Smart contract weakness classification (swc) registry,''
  \url{https://swcregistry.io/}.

\bibitem{defi_hack_labs}
\BIBentryALTinterwordspacing
SunWeb3Sec, ``Defihacklabs: Reproduce defi hacked incidents using foundry,''
  \url{https://github.com/SunWeb3Sec/DeFiHackLabs}. [Online]. Available:
  \url{https://github.com/SunWeb3Sec/DeFiHackLabs}
\BIBentrySTDinterwordspacing

\bibitem{si2025excessive}
W.~M. Si, M.~Li, M.~Backes, and Y.~Zhang, ``{Excessive Reasoning Attack on
  Reasoning LLMs},'' \emph{arXiv preprint arXiv:2506.14374}, 2025.

\bibitem{ghaleb2020effective}
A.~Ghaleb and K.~Pattabiraman, ``{How Effective are Smart Contract Analysis
  Tools? Evaluating Smart Contract Static Analysis Tools using Bug
  Injection},'' in \emph{Proceedings of the 29th ACM SIGSOFT International
  Symposium on Software Testing and Analysis}, 2020, pp. 415--427.

\bibitem{augusto2024xchainwatcher}
A.~Augusto, R.~Belchior, J.~Pfannschmidt, A.~Vasconcelos, and M.~Correia,
  ``{XChainWatcher: Monitoring and Identifying Attacks in Cross-Chain
  Bridges},'' \emph{arXiv preprint arXiv:2410.02029}, 2024.

\bibitem{CertiK2022CrossChainVulnerabilities}
{CertiK}, ``{Cross-Chain Vulnerabilities \& Bridge Exploits in 2022},''
  \url{https://www.certik.com/resources/blog/cross-chain-vulnerabilities-and-bridge-exploits-in-2022}.

\bibitem{Immunefi2023CommonCrossChainBridgeVulns}
{Immunefi}, ``{Common Cross-Chain Bridge Vulnerabilities},''
  \url{https://medium.com/immunefi/common-cross-chain-bridge-vulnerabilities-d8c161ffaf8f}.

\bibitem{sun2024gptscan}
Y.~Sun, D.~Wu, Y.~Xue, H.~Liu, H.~Wang, Z.~Xu, X.~Xie, and Y.~Liu, ``{GPTScan:
  Detecting Logic Vulnerabilities in Smart Contracts by Combining GPT with
  Program Analysis},'' in \emph{Proceedings of the IEEE/ACM 46th International
  Conference on Software Engineering}, 2024, pp. 1--13.

\bibitem{luu2016making}
L.~Luu, D.-H. Chu, H.~Olickel, P.~Saxena, and A.~Hobor, ``{Making Smart
  Contracts Smarter},'' in \emph{Proceedings of the 2016 ACM SIGSAC Conference
  on Computer and Communications Security}, 2016, pp. 254--269.

\bibitem{torres2018osiris}
C.~F. Torres, J.~Sch{\"u}tte, and R.~State, ``{Osiris: Hunting for Integer Bugs
  in Ethereum Smart Contracts},'' in \emph{Proceedings of the 34th Annual
  Computer Security Applications Conference}, 2018, pp. 664--676.

\bibitem{mueller2017framework}
B.~Mueller, ``{A Framework for Bug Hunting on the Ethereum Blockchain},''
  \emph{ConsenSys/mythril}, 2017.

\bibitem{mossberg2019manticore}
M.~Mossberg, F.~Manzano, E.~Hennenfent, A.~Groce, G.~Grieco, J.~Feist,
  T.~Brunson, and A.~Dinaburg, ``{Manticore: A User-Friendly Symbolic Execution
  Framework for Binaries and Smart Contracts},'' in \emph{2019 34th IEEE/ACM
  International Conference on Automated Software Engineering (ASE)}.\hskip 1em
  plus 0.5em minus 0.4em\relax IEEE, 2019, pp. 1186--1189.

\bibitem{nguyen2020sfuzz}
T.~D. Nguyen, L.~H. Pham, J.~Sun, Y.~Lin, and Q.~T. Minh, ``{sFuzz: An
  Efficient Adaptive Fuzzer for Solidity Smart Contracts},'' in
  \emph{Proceedings of the ACM/IEEE 42nd International Conference on Software
  Engineering}, 2020, pp. 778--788.

\bibitem{jiang2018contractfuzzer}
B.~Jiang, Y.~Liu, and W.~K. Chan, ``{ContractFuzzer: Fuzzing Smart Contracts
  for Vulnerability Detection},'' in \emph{Proceedings of the 33rd ACM/IEEE
  International Conference on Automated Software Engineering}, 2018, pp.
  259--269.

\bibitem{wu2025detecting}
H.~Wu, H.~Wang, S.~Li, Y.~Wu, M.~Fan, Y.~Zhao, and T.~Liu, ``{Detecting State
  Manipulation Vulnerabilities in Smart Contracts Using LLM and Static
  Analysis},'' \emph{arXiv preprint arXiv:2506.08561}, 2025.

\end{thebibliography}

\end{document}